\documentclass[fleqn,usenatbib]{mnras}

\usepackage{newtxtext,newtxmath}
\usepackage[T1]{fontenc}
\usepackage{ae,aecompl}

\usepackage{graphicx}	% Including figure files
\usepackage{amsmath}	% Advanced maths commands

\usepackage[flushleft]{threeparttable}

\usepackage{afterpage}
\usepackage{xspace}
\usepackage[dvipsnames]{xcolor}
\usepackage[normalem]{ulem}
\newcommand{\chandra}{\textsl{Chandra}\xspace}
\newcommand{\so}{4U~1700$-$37\xspace}
\newcommand{\redtext}[1]{{\color{red} #1}}

\title[\chandra view of \so in eclipse]{High resolution X-ray spectroscopy of Supergiant HMXB 4U1700$-$37 during the compact object eclipse.}

\author[]{M. Mart\'{i}nez-Chicharro$^{1}$\thanks{E-mail: maria.chicharro@ua.es},
V. Grinberg$^{2}$,
J.M. Torrej\'{o}n$^{1}$,
N. Schulz$^{3}$,
L. Oskinova$^{4}$,
\newauthor
M. Nowak$^{5}$,
F. F\"urst$^{6}$,
N. Hell$^{7}$ 
and R. Hainich$^{4}$
\\
$^{1}$Instituto Universitario de F\'isica Aplicada a las Ciencias y las Tecnolog\'ias (IUFACyT), Universidad de Alicante, E-03690 Alicante, Spain\\
$^{2}$Institut f\"ur Astronomie und Astrophysik, Universit\"at T\"ubingen,
       Sand 1, 72076 T\"ubingen, Germany\\
$^{3}$MIT:  Kavli Institute for Astrophysics and Space Research, Massachusetts Institute of Technology, Cambridge, MA 02139\\
$^{4}$Postdam Institut \"ur Physik und Astronomie, Universit\"at Potsdam, Karl-Liebknecht-Str. 24/25, D-14476 Potsdam, Germany\\
$^{5}$Department of Physics, Washington University, St. Louis, USA\\
$^{6}$ESA/ESAC: European Space Astronomy Centre (ESA/ESAC), Science Operations Department, Villanueva de la Ca\~{n}ada (Madrid), Spain\\
$^{7}$LLNL:Lawrence Livermore National Laboratory, 7000 East Ave., Livermore, CA 94550, USA 
}

\date{Accepted XXX. Received YYY; in original form ZZZ}

\pubyear{2019}

\begin{document}
\label{firstpage}
\pagerange{\pageref{firstpage}--\pageref{lastpage}}
\maketitle

\begin{abstract}
%This is a simple template for authors to write new MNRAS papers.
%The abstract should briefly describe the aims, methods, and main results of the paper.
%It should be a single paragraph not more than 250 words (200 words for Letters).
%No references should appear in the abstract.
We present an analysis of the first observation of the iconic High Mass X-ray Binary \so with the \chandra High Energy Transmission Gratings during an X-ray eclipse. The goal of the observation was to study the structure/physical conditions in the clumpy stellar wind through high resolution spectroscopy. We find that: a) emission line brightness from K shell transitions, corresponding to near neutral species, directly correlates with continuum illumination. However, these lines do not greatly diminish during eclipse. This is readily explained if fluorescence K$\alpha$ emission comes from the bulk of the wind. b) The highly ionised \ion{Fe}{xxv}{} and \ion{Fe}{xxvi}{} Ly$\alpha$ diminish during eclipse. Thus, they must be produced in the vicinity of the compact object where $\log \xi >3$. c) to describe the emission line spectrum, the sum of two self consistent photo ionisation models with low ionisation ($\log \xi\sim -1$) and high ionisation ($\log \xi\sim 2.4$) is required. From their emission measures, the clump-to-interclump density ratio can be estimated to be $n_c/n_i\sim 300$. To fit the complex He-like \ion{Si}{xiii}{} profile, the plasma requires a broadening with $v_{\rm bulk}\sim 840$ km s$^{-1}$. Reproducing the observed $r\approx f$ line fluxes requires the addition of a third collisionally ionised plasma. d) Emission lines widths appear unresolved at the \textsc{hetg} gratings resolution with exception of Silicon. There is no clear radial segregation between (quasi)neutral and ionised species, consistent with cold wind clumps interspersed in a hot rarefied interclump medium.

\end{abstract}


%%%%%%%%%%%%%%%%%%%%%%%%%%%%%%%%%%%%%%%%%%%%%%%%%%

%%%%%%%%%%%%%%%%% BODY OF PAPER %%%%%%%%%%%%%%%%%%

\section{Introduction}

Massive stars ($M_{\rm i}\gtrsim$\,8\,$M_\odot$) are crucial cosmic engines. Their strong radiation driven stellar winds $($responsible for the mass loss of the star$)$  and their final supernova explosion provide a significant input of matter, mechanical energy and momentum into interstellar space, triggering star formation and enriching the interstellar medium with heavy elements that, ultimately, enable Earth-like rocky planets and life. Yet the structure and properties of massive star winds are still poorly understood. The structured and clumped wind paradigm is well established \citep[e.g.,][]{1997A&A...322..878F}. The wind acceleration mechanism is intrinsically unstable and soon develops high density areas (clumps) separated by more rarefied sections (interclump medium), as compared with the original radially smoothly varying wind. However, there are still serious discrepancies between the different model predictions as well as between the models and observations \citep[e.g.,][]{2012MNRAS.421.2820O}. The inner parts of OB-star winds ($a< 1.25R_{*}$)  are inhomogeneous and clumped \citep{2006A&A...454..625P, 2015ApJ...810..102T, Sundqvist_2018a}, and their complex properties are poorly understood. 

In high mass X-ray binaries (HMXBs) with supergiant donors, a compact object (neutron star or black hole) is on a relatively close orbit, deeply embedded into the wind of its donor star. The accretion of matter from the stellar wind powers strong X-ray radiation that illuminates nearby wind regions. This radiation excites transitions in the stellar wind that can be used as a unique diagnostic of wind properties \citep{2017SSRv..212...59M}. 

%%%%%%%%%%PARAMETERS SYSTEM HAINICH%%%%%%%%%%%%
\begin{table}
{\def\arraystretch{1.3}
%\begin{center}
\centering
\caption{Parameters of the 4U1700$-$37 system (after \citealt{2020A&A...634A..49H}).
\label{Parametes system }}
\begin{threeparttable}
\label{tab:Parameters system 2}

\begin{tabular}{lcc}
\hline

Parameter	&	Symbol	&	Value		\\
\hline							
Distance 	&	$d$	& $1.7^{+0.3}_{-0.2}$\,kpc	\\

O star Temperature  	&	$T_{*}$ &$35^{+2}_{-3}$\,kK\\
O star radius 	&	$R_{*}$ &$19^{+5}_{-6}$ ($R_{\odot}$) 		\\
Clumping factor   & $D$ & $20^{+50}_{-15}$ \\
O star mass	&	$M_{\mathrm{spec}}$	&	$34^{+100}_{-28}$ ($M_{\odot}$)\\
Spectral type 	&		& O6Iafpe	\\

Wind terminal velocity	&	$v_{\infty}$ & $1900^{+100}_{-100}$	(km\,s$^{-1}$)	\\
Wind mass-loss rate 	&	log ${\dot{M}}$ & $-5.6^{+0.2}_{-0.3}$ ($M_{\odot} \mathrm{yr}^{-1})$		\\
Average orbital distance 	& $a_{2}$ & $1.6^{+0.5}_{-0.4}$ ($R_{*}$)		 	\\
Wind velocity law  & $\beta$ & $2^{+1}_{-1}$ \\
Interstellar reddening & $E(B-V)$ & $0.50\pm0.01$ \\

Orbital period 	& $P_{\rm orb}$ & $3.411660\pm0.000004$	(d)	\tnote{a}\\

\hline
\end{tabular}																	
\begin{tablenotes}
\item[a] From \cite{2016MNRAS.461..816I}.
\end{tablenotes}	
																		
\end{threeparttable}	

%\end{center}																		
}																		
\end{table}

In \so, discovered with the \textsl{Uhuru} satellite \citep{1973ApJ...181L..43J},  the stellar wind of the O6Ia supergiant HD~153919 (= V884 Sco), the earliest donor in any Galactic HMXB, is ionised by the strong
persistent X-rays from a compact object companion. The latter is on a close orbit deep in the innermost region of the donor star's wind; the best estimates for the parameters of the \so system are summarized in Table~\ref{tab:Parameters system 2}. The nature of the compact object is not fully clear yet due to the lack of coherent pulsations at X-rays or any other wavelengths. However, \citet{1999A&A...349..873R} show that the 2--200\,keV spectrum of 4U 1700$-$37 is different from those commonly observed for black hole candidates, such as Cyg X-1, but qualitatively similar to those of accreting neutron stars. They explain the lack of pulsations as due to either a weak magnetic field or an alignment of the magnetic field with the spin axis.
The neutron star nature of the compact object has been proposed  with indirect evidence based on the X-ray spectra \citep{2016ApJ...821...23S} and the X-ray colour-colour behaviour \citep{2003ApJ...592..516B}. We have provided further strong evidence for the neutron star nature based on the spectral behavior in quiescence and during a flare, as seen with \textsl{Chandra} \citep{2018MNRAS.473L..74M}. Additionally, the possible detection on a cyclotron scattering resonance feature has been proposed recently by \citet{Bala_2020a}, that would be a direct evidence of the presence of a strong magnetic field of the order of $\sim10^{12}$\,G and thus a neutron star compact object.

\so is located at a distance of only $\sim$1.7\,kpc from Earth \citep{2018yCat..36160037B}\footnote{Distances based on the Gaia DR2 measurements \citep{2018A&A...616A...1G}, calculated by means of a Bayesian approach assuming an exponentially decreasing space density with the distance.} and is fairly bright ($8-45\times 10^{-10}$\,erg\,s$^{-1}$\,cm$^{-2}$,  \citealt{2018MNRAS.473L..74M}). It has thus been observed with virtually all X-ray telescopes since the dawn of X-ray astronomy. It shows strong flaring activity, with flux increases by a factor of ten and above \citep{Kuulkers_2007a}. These periods last $\sim$an hour; the lightcurves during these periods show strong flickering \citep{2003ApJ...592..516B}. There is no consensus about the origin of these flares yet; one possible explanation is an accretion episode from \redtext{t}he magnetotail of the neutron star \citep{1981A&A....94..323B}.

Studies of \so at high spectral resolution in X-rays are scarce. \citet{2005A&A...432..999V} reported results from observations of \so with \textsl{XMM-Newton} at several orbital phases and presented a thorough study with EPIC-MOS at CCD resolution. The spectra show prominent Fe lines and a number of other species in the low energy band. Unfortunately, the reflection grating spectrometer (RGS) spectrum had a poor S/N ratio that prevented any emission line analysis. 
 
\begin{figure}
\includegraphics[width=1.\columnwidth]{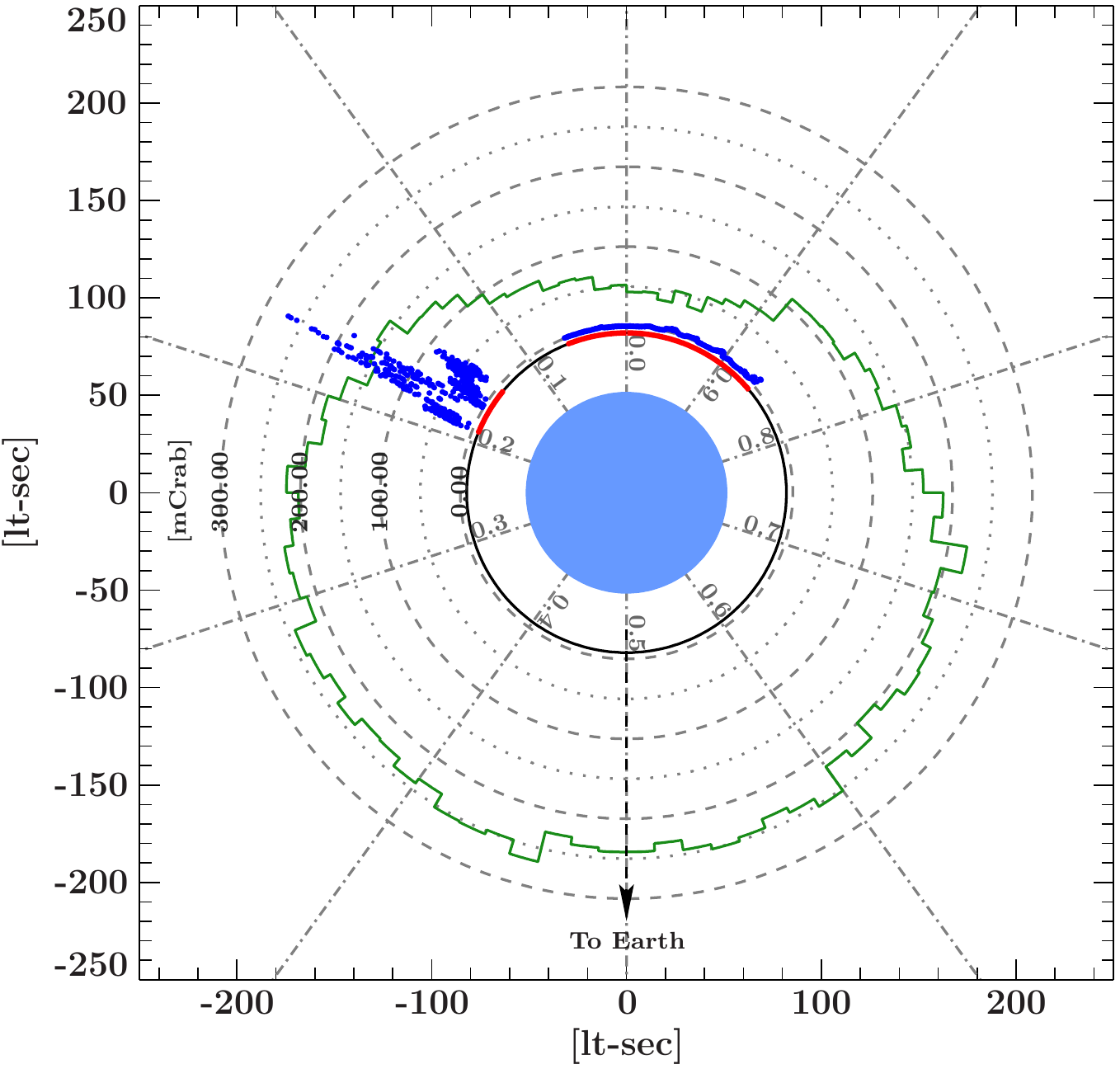}

\caption{To scale sketch of the \so system.  We use the folded \textsl{Swift}-BAT light curve (shown in green) as an indicator of general system brightness and to trace the X-ray eclipse. The Swift-BAT covers 
the energy range of 15-150\,keV that is little affected by absorption.The red line shows the phases of the \chandra  ObsID~18951 (this paper; $\phi_{\mathrm{orb}} \approx 0.85-0.05$) and ObsID~17630 \citep{2018MNRAS.473L..74M}. Blue points represent count rates during \chandra observations n the energy range of 0.8-7.7\,KeV 
in counts s$^{-1}$. Short term variability (specially an outburst at the end of ObsID 17630) is visible in the 
Chandra observations, but it is smeared out in the long term \textsl{Swift}-BAT light curve. }
\label{fig:orbital}
\end{figure}

\citet{2003ApJ...592..516B} performed the first high$-$resolution analysis with \chandra-\textsc{hetg} at orbital phases $\phi_{\mathrm{orb}} \approx 0.65-0.80$, before eclipse (Fig. \ref{fig:orbital}).  This study looks at line variability as the source flares; the plethora of emission lines detected can be used for plasma diagnostics. \citet{2020A&A...634A..49H} analysed \textit{Hubble Space Telescope} (\textit{HST}) UV and \textit{Fiber-fed Extended Range Optical Spectrograph} (FEROS) optical spectra of \so using the \textsl{PoWR} stellar atmosphere code\footnote{\url{http://www.astro.physik.uni-potsdam.de/~wrh/PoWR/powrgrid1.php}}. The parameters so measured are presented in Table \ref{tab:Parameters system 2}.

%%Orbital PLOT %%%%%%%%%%%%%%%%%%%%%%
%%%%%%%%%%%%%%%%%%%%% END Orbital plot  %%%%%%%%%%%%%%%

Studies of HMXBs during the compact object eclipse have been fundamental to probe the emitting plasma properties through emission line analysis \citep[i.e.][for Vela X-1 using \textsl{ASCA}]{1994ApJ...436L...1N,1999ApJ...525..921S} showing the coexistence of different ionisation states within the stellar wind volume. Comparison of eclipse to out-of-eclipse spectra also allows to probe emitting structures within HMXBs \citep[][]{2019ApJS..243...29A}.  
In this paper, we present the first high$-$resolution spectrum of \so taken during the X-ray eclipse. During the eclipse, the suppression of the continuum (typically to $\sim$ a tenth of the uneclipsed flux, \citep[][]{2015A&A...576A.108G, 2019ApJS..243...29A} emitted by the compact object allows to observe the emission line spectrum from the back illuminated stellar wind of the O6Ia star HD153919 with unprecedented detail. 
The paper is structured as follows: in Section \ref{sec:obs}, we present the observational details. In section \ref{sec:lc}, we analyse the IR-optical and X-ray light curves.  In Section \ref{sec:analysis}, we present a description of the spectral analysis and results. 
Finally, in Sections \ref{sec:disc} and \ref{sec:conc}, we discuss the parameters obtained in order to understand the origin of emission lines and present our conclusions. 

%\section{Observation}
\section{Observations}
\label{sec:obs}

We performed a pointed observation of \so with \chandra on 29 June 2017. The High Energy Transmission Grating Spectrometer (\textsc{hetg}; \citealt{2005PASP..117.1144C}) aboard the \chandra X-Ray Observatory \citep{2002PASP..114....1W} acquired data uninterrupted for 58\,ks. We scheduled the observation to coincide with the eclipse, at $\phi _{\rm orb}$= 0.848$-$0.044, so that we could investigate the excited emission lines with the highest line-to-continuum contrast. There are two sets of gratings available, the High Energy Grating (\textsc{heg}) which offers a  resolution of 0.011 \AA\ in the bandpass of about 1.5 to 16 \AA,
and the Medium Energy Grating (\textsc{meg}) which offers a resolution of 0.021 \AA\ in the the bandpass of about 1.8 to 31 \AA. Our observations provided significant data in the range between 1.6 and 10$-$20 \AA.

The spectra were reduced and response (\texttt{arf} and \texttt{rmf}) files were generated using standard procedures with the \textsc{ciao} software (v4.11, CalDB 4.7.8). First dispersion orders ($m=\pm 1$) from \textsc{heg} and \textsc{meg} were extracted and combined into a single spectrum. The peak source count rate
%vg: it's not flux if it is in counts :)
both in \textsc{heg} and  in \textsc{meg} gratings is 0.3 counts\,s$^{-1}$, which is much lower than the level at which pileup effects start to become important in the grating spectra\footnote {See \emph{The Chandra ABC Guide to Pileup}, v.2.2, 
\url{https://cxc.harvard.edu/ciao/download/doc/pileup_abc.pdf}}. We also extracted the 0th order \textsc{acis-i} spectrum, but found out it was piled-up, so that we are not using it in this paper. The spectral analysis was performed with the Interactive Spectral Interpretation System (\textsc{isis}) v 1.6.1-24 \citep{2000ASPC..216..591H}.

 As an out-of-eclipse comparison spectrum we use the ObsID 17630 data analysed in \citet{2018MNRAS.473L..74M}, acquired after egress ($\phi=0.16$). An sketch of the \so system is presented in Fig. \ref{fig:orbital}. The two ObsIDs analyzed in this paper are marked in red and their lightcurves shown in blue. 

%%%%TABLE OBSERVATION %%%%%%%%%%%%%
%%%%%%%%%%%%%%%%%%%%%%%%%%%%%%%%%%%
\begin{table}
	\centering
	\caption{Observations journal}
\begin{threeparttable}
	\label{tab:obs}
	\begin{tabular}{lccr} % four columns, alignment for each
		\hline
		ObsID & Date & $t_{\rm exp}$ & $\phi _{\rm orb}$ \\
		\hline
		18951 &  2017-06-29 22:02:38 & 57.8 ks &  0.848$-$0.044\tnote{a}\\
		17630 &  2015-02-22 03:11:16 & 14.3 ks &  0.16\tnote{a,b}\\
		\hline
	\end{tabular}
\begin{tablenotes}
\item[a] Mid eclipse time $T_{0}=49149.412\pm 0.006$ MJD, orbital period
$P=3.411 660\pm0.000 004$ d \citep{2016MNRAS.461..816I}
\item[b] Out of eclipse. 
\end{tablenotes}
\end{threeparttable}
\end{table}
%%%%%%%%%%%%%%%%%%%END TABLE OBSERVATION %%%%%%%%%%%%%%%%%

\section{Light curves}
\label{sec:lc}

%%LIGHTCURVE PLOT %%%%%%%%%%%%%%%%%%%%%%
\begin{figure}

\includegraphics[width=1.0\columnwidth]{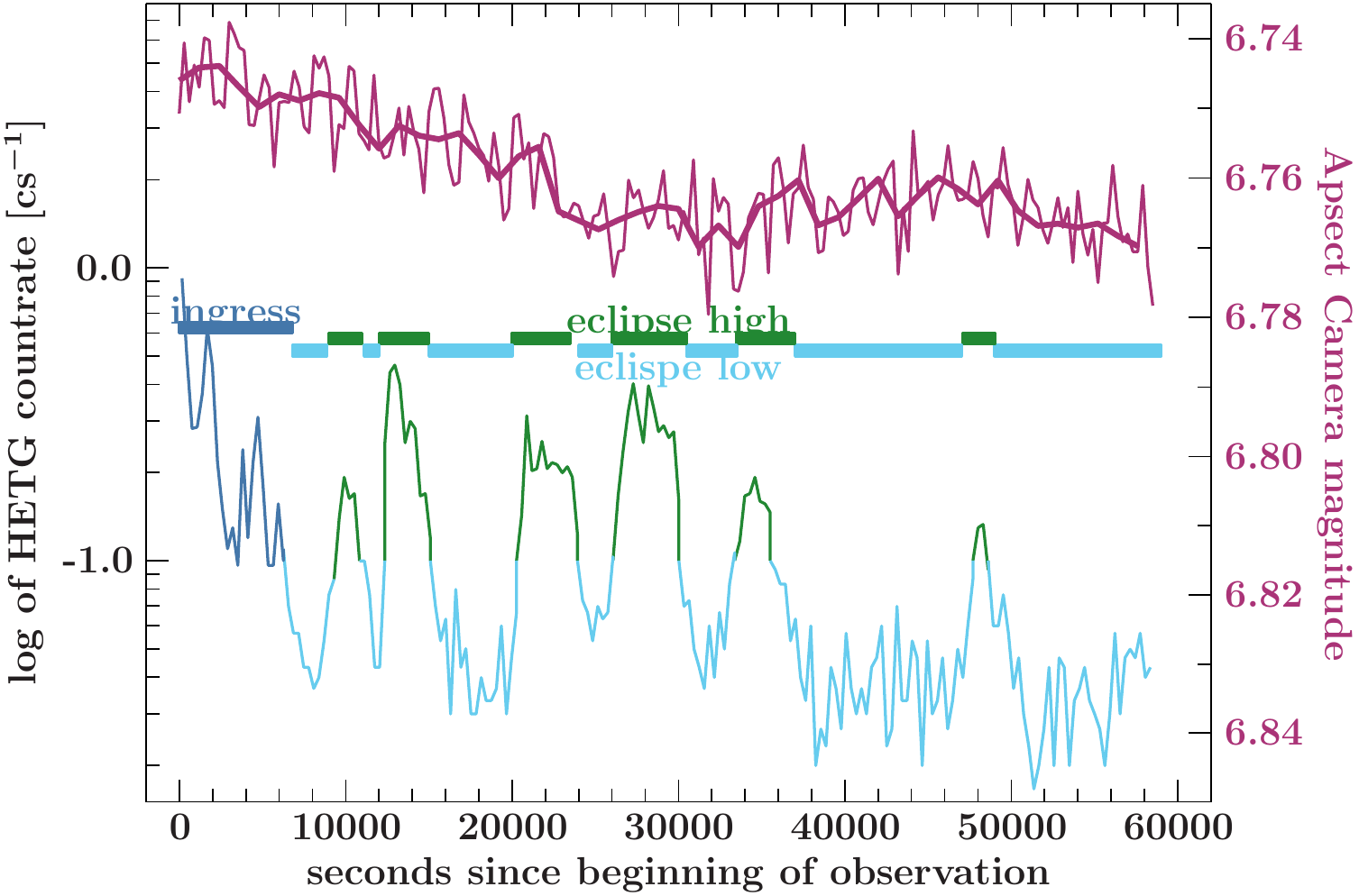}

\caption{\chandra Aspect Camera (4000-8000\,{\AA}, purple) and \textsc{HETG}  (1.6 - 15 (\AA), green, blue) light curves of \so, during ObsID 18951. We divide the X-ray light curve into three separate extractions: 1) Ingress (dark blue), leading into eclipse, 2) eclipse high (green) (> 0.1 counts s$^{-1}$) and 3) eclipse low (blue).}
\label{fig:lcs}
\end{figure}
%%%%%%%%%%%%%%%%%%%%% END PLOT LIGHT CURVE %%%%%%%%%%%%%%%

In Fig.~\ref{fig:lcs} we show the \textsc{hetg} X-ray light curve (bottom, black) for the combined \textsc{heg} and \textsc{meg} first-order spectra and the strictly simultaneous optical-NIR light curve (top, red) acquired with the on board Aspect Camera, both in 300\,s bins. The Aspect Camera is a broadband detector in approximately 4000-8000\,{\AA} range.  The magnitudes reported are in the photometric system specific to this camera \citep{2010ApJS..188..473N}. The apparent pulsation seen in the optical light curve is caused by the telescope dithering pattern that has nominal periods of 1\,ks and $\sim0.7$\,ks in Y and Z-directions, respectively. Thus we also over plot the 1200 s running average. 

Both light curves show a remarkable behaviour during the observation. After the eclipse ingress, the X-ray light curve displays flares at the beginning of the eclipse, when the neutron star is already hidden from direct view by the donor star. Such flares are well known in \so  and have been discussed in \cite{2016MNRAS.461..816I}. Later, during the second half of observation, the light curve shows less pronounced variability. At the same time, the optical light curve seems to display a dimming (magnitude increase) until the end of the X-ray flaring when the dimming stops and the lightcurve reaches an average value consistent with the donor star photometric $B$ magnitude. 

In order to perform the X-ray spectral analysis we extracted data, separately, from the eclipse ingress (from $t=0$ to $t\sim6$\,ks, counting from the beginning of the observation) and the eclipse. We subdivide the eclipse further into eclipse low ($<0.1$\,counts s$^{-1}$) and eclipse high regimes ($>0.1$\,counts s$^{-1}$). Since our goal is to explore the back illuminated stellar wind, we will concentrate on the last two. These two regimes are indicated in Fig.~\ref{fig:lcs}, respectively, and will be used in the next sections.

\section{Spectral analysis}
\label{sec:analysis}
In the following analysis, we fit the models to unbinned data We use the C-statistic \citep{1979ApJ...228..939C}, which is appropriate when the bins have few counts (typically less than 20 counts per bin during eclipse), and the \texttt{subplex} fitting method. 

\subsection{Continuum modeling}

The suppression of the continuum during the eclipse allows us to analyze the emission lines excited in the stellar wind with a high contrast. At the same time, it complicates also the definition of the continuum because the spectrum is dominated by strong emission lines that cannot be ignored during the continuum modelling. We thus proceed as follows to model the continuum in the $1.6-20$\,{\AA} range: first, to find the lines, we use a blind line search where we start with a "test" phenomenological continuum (a simple power law, with partial covering, as described by Eq. \ref{eq:abs}) and then loop through the data adding one line at a time. The line search is over narrow energy/wavelength bands scanning the whole range of data with any previous lines and continuum fixed. Then we choose the line that changes the statistic the most and refit the continuum and all previous lines while limiting the energy of the previously found lines to narrow ranges around the expected values. We use this line identification approach to compare with the next part of the study, in which we use the Bayesian Blocks method (see Section \ref{sec:emission_line_spectrum}). 

For a blind line search as we have performed, determining the statistical significance of
a given line that takes into account both the change in fit statistic and the multiplicity of the 
line  searches is still a matter of research \citep[i.e.][]{2018arXiv181002207B}. Even these efforts do not take into account
systematic uncertainties \citep[e.g., from using a phenomenological continuum model, as well as 
uncertainties from the detector responses, ][]{2014ApJ...794...97X}. For our purposes,
we have kept track of the statistical order in which the lines have been added and we have only
kept lines where the change in fit statistic from adding a given line was $\Delta C^{2}\geq 0.005$.

Once we complete this procedure, we change the continuum model to a more physically motivated one, namely the continuum model used to describe ObsID 17630 performed out of eclipse (\citealt{2018MNRAS.473L..74M}, see also Fig.~\ref{fig:orbital}). The spectra are described by the Bulk Motion Comptonization model or \texttt{bmc} \citep{1997ApJ...487..834T}. In this model, soft photons with a characteristic color temperature $kT_{\rm col}$, are upscattered to high energies. The efficiency of the Comptonization is measured by the spectral parameter $\alpha$ (higher efficiency for lower values of $\alpha$). 

This continuum is modified at low energies by a partial absorber modeled as:
\begin{equation}
\label{eq:abs}
    \texttt{abs}(E)=\epsilon\exp{(-\sigma(E)N_{\rm H,1})}+(1-\epsilon)\exp{(-\sigma(E)N_{\rm H,2})}
\end{equation}
using the model \texttt{TBnew} with cross sections by \citet{Verner_1996a} and \citet{Wilms_2000a} abundances. $\epsilon$ is the covering fraction by the local plus interstellar medium (ISM) absorber with total column density $N_{\rm H,1}=N_{\rm local}+N_{\rm ISM}$. Actually, the local absorption is graded, and even the scattered component will be seen through a variety of density columns, that we model here as a single one. $N_{\rm H,2}=N_{\rm ISM}$ describes absorption by the interstellar medium. In our fits, it has been fixed to the ISM value deduced from the optical and UV observations using the value for $E(B-V)$ from Table \ref{tab:Parameters system 2} and $N_{\rm H}=0.65E(B-V)$ as defined in \citet{2015ApJ...809...66V}. $N_{\rm H,1}$ is left to vary freely. 

\begin{figure*}

\includegraphics[width=1\columnwidth]{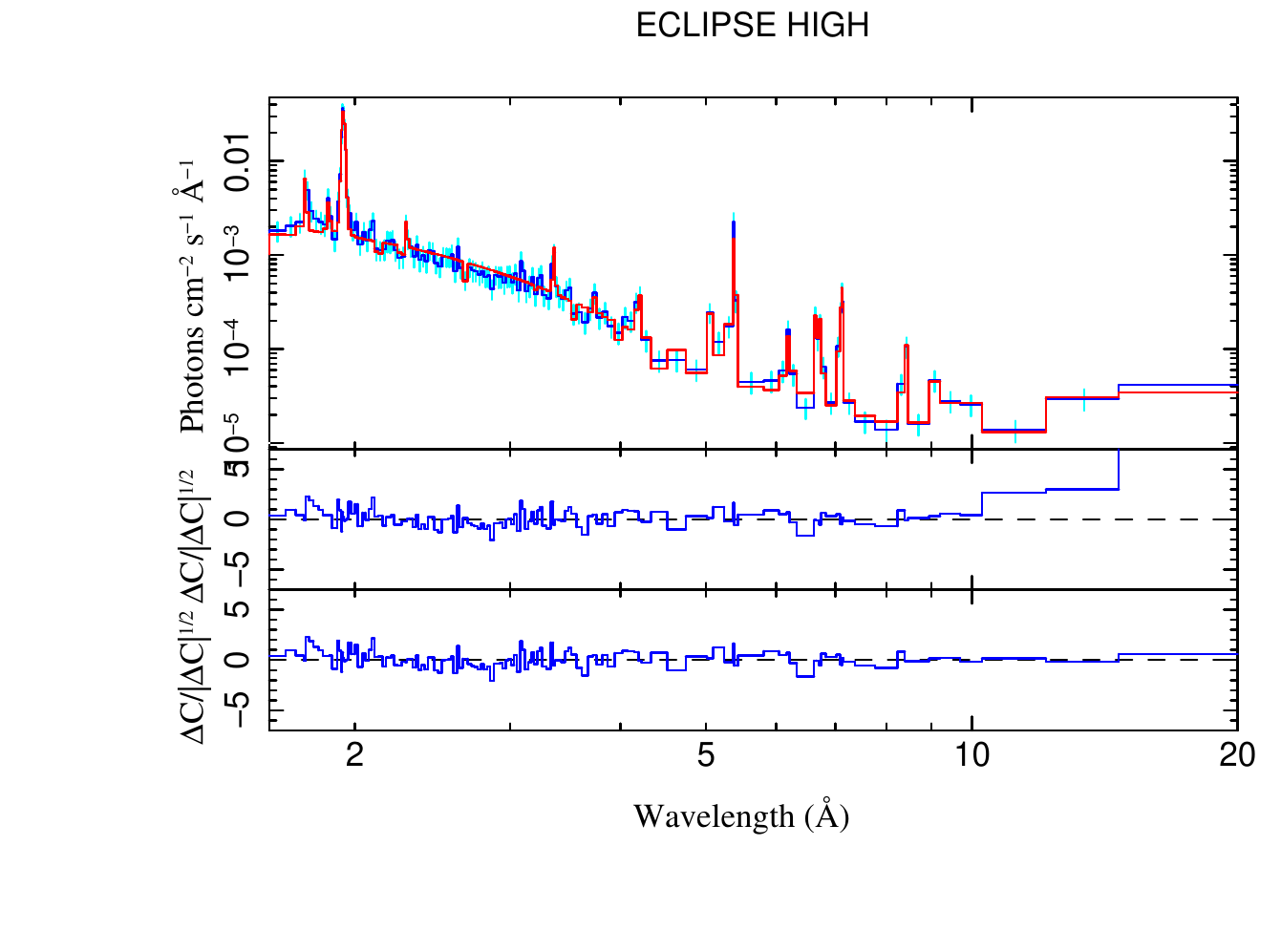}
\includegraphics[width=1\columnwidth]{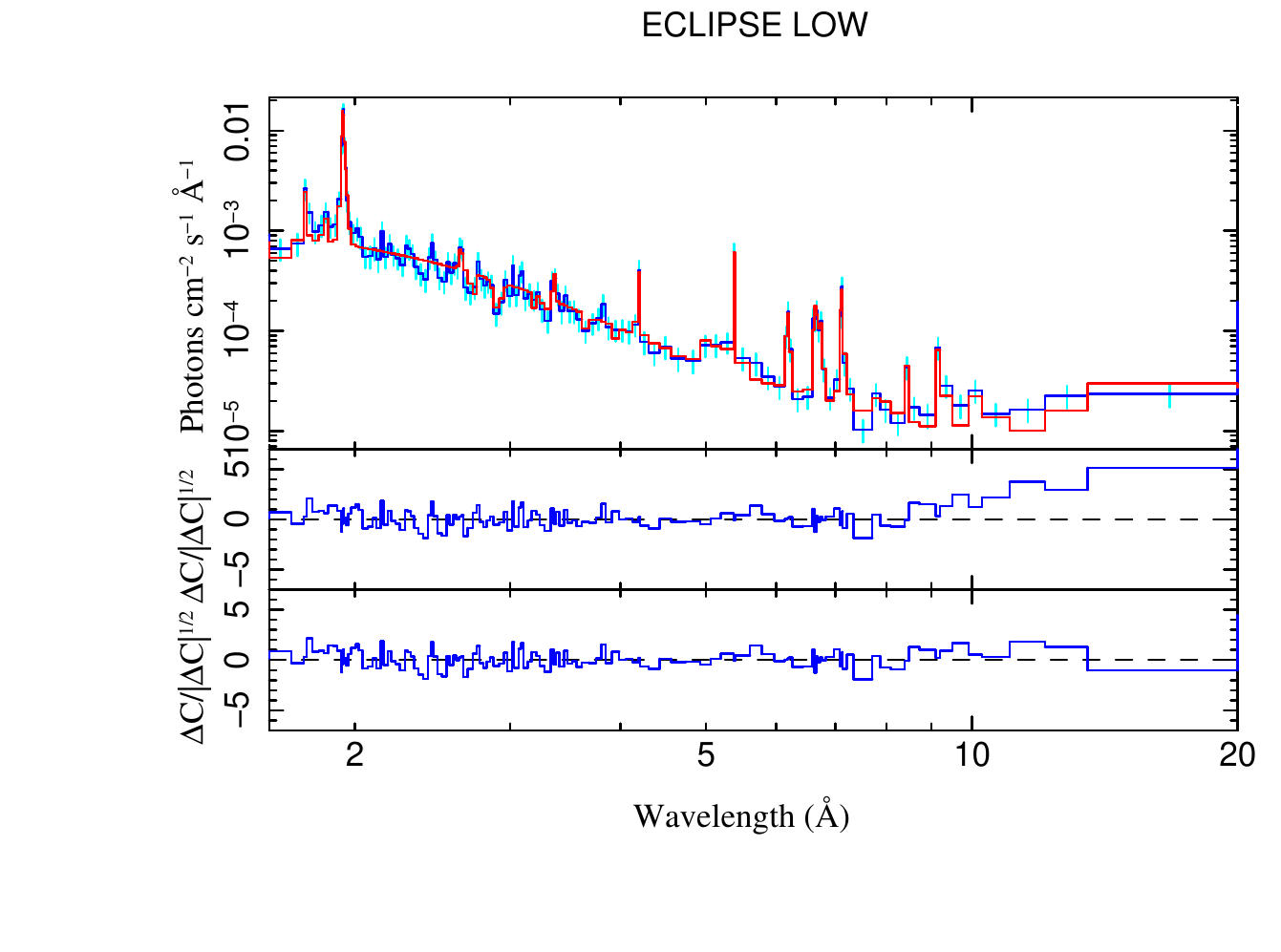}
\caption{\chandra spectrum from 1.65 to 20\,{\AA}, data (blue) and model fit (red), for 
eclipse high (left) and eclipse low (right) regimes.
Bottom panels show the residuals with (lowest panel) and without (middle panel) the addition of a blackbody at low energies.} 
\label{fig:HF_LF_Mike_H}
\end{figure*}

This model describes the spectrum well overall (Fig.~\ref{fig:HF_LF_Mike_H}), but residuals remain at low energies. This \textit{soft excess} is commonly seen in the spectra of HMXBs and its exact nature is still unclear \citep{2004ApJ...614..881H}. One possibility is that it is formed by unresolved Fe L emission lines grouping in this wavelength range \citep{2002ApJS..140..589B} but the lack of resolution prevents any further analysis in this respect. We thus empirically model it with a black body of $kT_{\rm bb}\sim 0.1$\,keV modified by its own absorption $N_{\rm H,3}$. The absorption column turns out to be compatible with the ISM value. The best fit parameters are presented in Table~\ref{tab:continuum_bcm} and the data (strongly rebinned for plotting purposes), the model (red) and the residuals are presented in Fig.~\ref{fig:HF_LF_Mike_H}. Although the spectra in and out of eclipse show differences (i.e. different photon indexes) owing, for example, to the energy dependence of the scattering \citep{2019ApJS..243...29A}, the obtained spectral parameters are in line with those deduced out of eclipse by \citet{2018MNRAS.473L..74M}, albeit with a different spectral index $\alpha$.

For a distance of 1.7 kpc, the \texttt{bmc} fluxes in Table~\ref{tab:continuum_bcm} correspond to X-ray luminosities of 
$L_{\rm X}=(0.5^{+0.2}_{-0.3})\times 10^{34}$\,erg\,s$^{-1}$ (eclipse low) and $L_{\rm X}=(2.5^{+0.1}_{-0.3})\times 10^{34}$\,erg\,s$^{-1}$ (eclipse high), in the $1.5-20$ \AA\ range.
Thus the average is a factor about 10 times lower than that observed out of eclipse during single \textit{XMM-Newton} observations \citep{2015A&A...576A.108G,2019ApJS..243...29A} and 20 times lower than during ObsID 17630 (quiescence) reported by \cite{2018MNRAS.473L..74M}. When comparing these results it must be taken into account that the source was intrinsically brighter during ObsID 17630. Indeed, the \textit{Swift-BAT} telescope count rate was $\approx 0.04$ c s$^{-1}$ which is twice that of ObsID18951 and the long term average. At the same time, they are two orders of magnitude higher that those displayed by O supergiants that are not in a binary system with a compact object \citep{2018A&A...620A..89N}.

\begin{table}
{\def\arraystretch{1.3}
\begin{center}
\caption{Model \texttt{bmc + bb } continuum parameters.}
\begin{threeparttable}
\label{tab:continuum_bcm}
\begin{tabular}{lcc}
\hline\hline
Parameter &  Eclipse Low & Eclipse High \\
\hline
&\texttt{bmc}&\\
$N_{\rm H,1}$ [10$^{22}$\,cm$^{-2}$]	& $21.5^{+2.2}_{-1.6}$		&	$21.1^{+1.3}_{-1.4}$	\\

$\epsilon$ 		&	$0.79^{+0.01}_{-0.02}$	&	$0.89^{+0.01}_{-0.01}$		\\
$N_{\rm H,2}$ [10$^{22}$\,cm$^{-2}$]		& $0.3$	&	$0.3^{+0.3}_{-0.2}$	\\

Norm 	[$\times 10^{-4}$]	&	  $1.9^{+0.1}_{-0.1}$	&	$21.1^{+0.9}_{-0.5}$		\\
$kT_{\rm col}$ [keV]		& $1.59^{+0.04}_{-0.01}$		&	$1.49^{+0.04}_{-0.04}$		\\
$\alpha$ 		&	$0.99^{+0.12}_{-0.08}$	&	$0.19^{+0.02}_{-0.01}$		\\
$f$		&	10	&	10		\\
Flux [$\times 10^{-11}$\,erg\,s$^{-1}$\,cm$^{-2}$]\tnote{a}		& 	$1.12^{+0.06}_{-0.06}$	&	$5.71^{+0.24}_{-0.13}$		\\
&&\\						
						
		&	\texttt{bb}	&			\\

$N_{\rm H,3}$ [10$^{22}$\,cm$^{-2}$]		& $0.3$	&	$0.3^{+0.1}_{-0.2}$	\\
Norm$_{\rm bb}$	[$\times 10^{-5}$]	&	$7^{+1}_{-1}$   	&	   	$15^{+4}_{-5}$	\\
$kT_{\rm bb}$ [keV]	&	$0.10^{+0.01}_{-0.01}$	&	$0.10^{+0.01}_{-0.01}$		\\

Flux	[$\times 10^{-11}$\,erg\,s$^{-1}$\,cm$^{-2}$]\tnote{a}		& $0.29^{+0.04}_{-0.04}$	&	$1.30^{+0.34}_{-0.43}$		\\
&&\\							
 C$^{2}_{\rm r}$ (d.o.f.)		&	0.77 (44)	&0.60	(38)		\\
\hline
	\end{tabular}
\begin{tablenotes}

\item[a]{Unabsorbed $1.5-20$ \AA\ flux}

\item[b]{Numbers without errors have been fixed at the quoted values}

\end{tablenotes}
\end{threeparttable}
\end{center}
}
\end{table}

\subsection{Emission line spectrum}
\label{sec:emission_line_spectrum}
Once the continuum has been modeled, we start the investigation of the emission line spectrum. Each line is modeled by adding a Gaussian component. 
In order to perform a blind search of spectral features we use the Bayesian blocks algorithm\footnote{As implemented in the SITAR package and included in the \texttt{isisscripts} \url{http:// www.sternwarte.uni-erlangen.de/isis/}.} \citep{2013arXiv1304.2818S}.  The Bayesian Blocks approach for line detection in high$-$resolution X$-$ray spectroscopy is introduced, discussed, and benchmarked against other methods in \citet{2007ApJ...669..830Y}. To assess the reliability of emission line detection, we list the parameter $\alpha_{\mathrm{sig}}$ that can be roughly related the significance of the feature detection as $P \approx 1 - \exp(-2\alpha_{\mathrm{sig}})$, as discussed previously in \citet{2017A&A...608A.143G}. We refer to \citet{2007ApJ...669..830Y} and \citet{2017A&A...608A.143G} for a more extensive discussion of the method.

The Bayesian Blocks based search is complemented by a conventional, manual approach, where we utilize our knowledge of expected line energies of a given element and ion, based on the line positions obtained from \texttt{AtomDB}\,v. \,3.0 \citep{Foster_2012a}.

%%IRON REGION PLOT
\begin{figure}

%ylog;
%\includegraphics[width=1\columnwidth]{Iron_region_x_10_to_yr.pdf}
%\includegraphics[width=1\columnwidth]{Iron_region_ylabel.pdf}
\includegraphics[width=1\columnwidth]{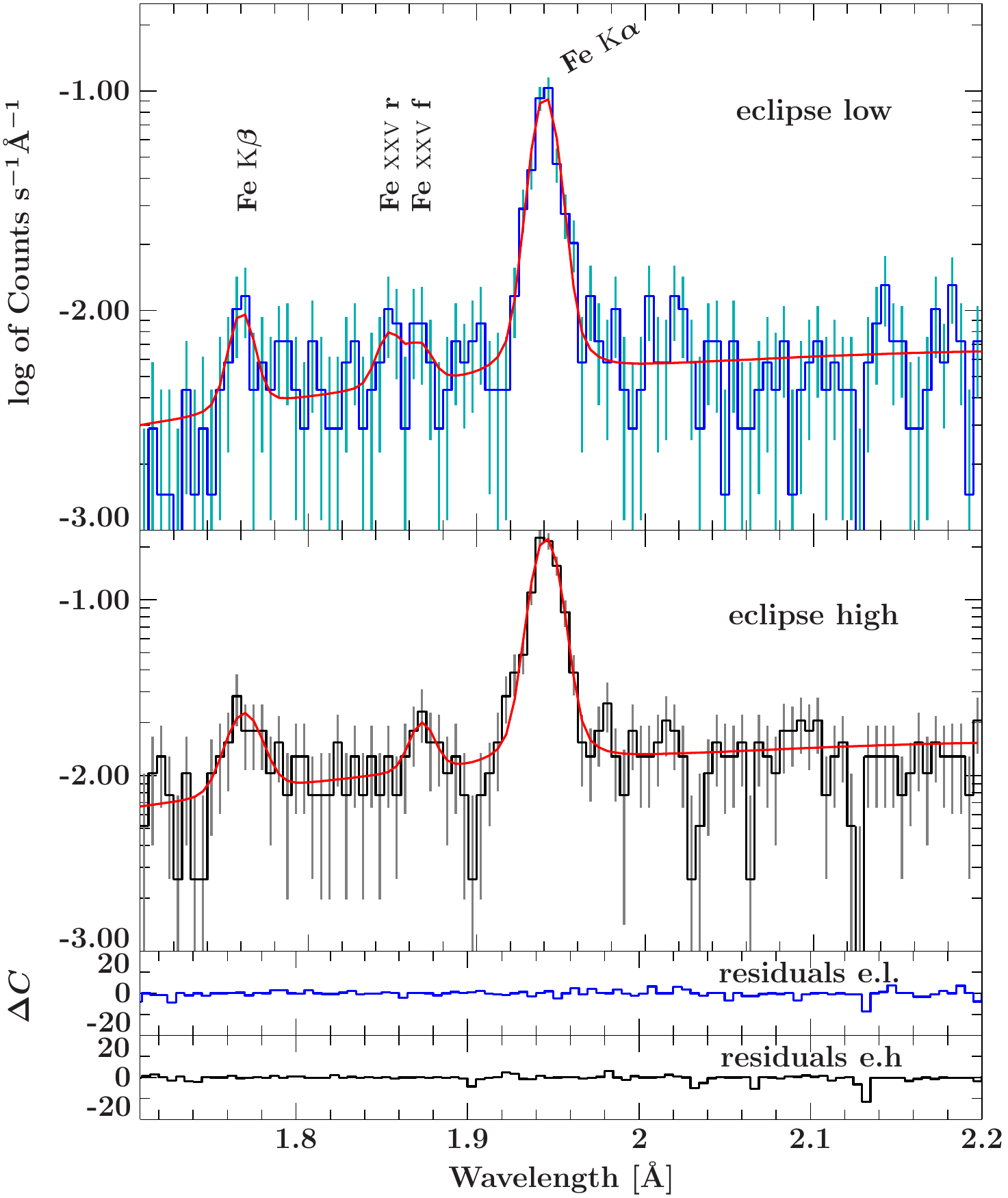}

\caption{\chandra emission lines from 1.65$-$2.2 \AA .The model is shown in red. The error bars in light blue and the data in dark blue (eclipse low). The error bars in gray and the data in black (eclipse high).}
\label{fig:Fe}
\end{figure}

We add, one by one, all the lines found by Bayesian Blocks, fitting the continuum again every time we add a new line. When no more lines are identified by the Bayesian Blocks, we switch to the manual approach explained above. The width of the lines is restricted to 0.005--0.1\,{\AA}. The position of the lines is restricted to within $\lambda_{0}\pm 0.01$\,{\AA}, where $\lambda_{0}$ is the laboratory wavelength. The errors have been calculated after restricting the parameters so their actual values could exceed the quoted errors for the weakest lines. 

To reduce the free parameters during the fits, we further link the wavelength of the $fir$ components of the He-like triplets to their theoretical differences and fit one line shift for the whole triplet. The intercombination ($i$) lines of the  He-like ions typically consists of two transitions($i1$, $i2$). These two transitions are unresolved in our spectra and we thus use their average wavelength, weighted by a factor 1:1\footnote{The statistical weights are $g_y = 2*1+1=3$ and $g_x = 2*2+1=5$, yielding 3:5 and not 1:1. $y$ is an electric dipole (E1) while $x$ is a magnetic quadrupole (M2) transition.  For the He-like triplet, the line ratios depend strongly on the excitation mechanism. For photo-excitation from the ground, it is not expected to see the M2 transition at all. For collisional excitation from the ground, the cross sections of these two upper levels would be more important than the statistical weights. However, with strong UV fields present or low-temperature, high density plasmas, the upper levels of $x$ and $y$ can be excited from 
the upper level 1s2s ${}^3S_1$ of the forbidden line ($f$ or $z$) instead of 
from the ground, while still preferentially decaying to the ground; the 
transition from the upper level of $z$ to either of the upper levels of $x$ 
and $y$ are E1 transitions, i.e., in this case the statistical weights 
would apply again. However, for low-Z elements the wavelength difference between $x$ and $y$ is fairly small, so the approximation adopted here is fairly good.}. For He-like \ion{Si}{XIII}~$i$, for example, we so obtain 6.686\,{\AA} using the wavelength of individual transitions as obtained from \texttt{AtomDB}. The lines of the Ly series of H-like ions typically also consists  of two strong transitions. These two transitions are unresolved in our spectra and we thus use their average wavelength, weighted by a factor 2:1 according to their statistical weight. For H-like \ion{Si}{vi} Ly$\alpha$, for example, we so obtain 6.1821\,{\AA} using the wavelength of individual transitions as obtained from \texttt{AtomDB}.

\begin{figure*}

\includegraphics[width=0.99\columnwidth]{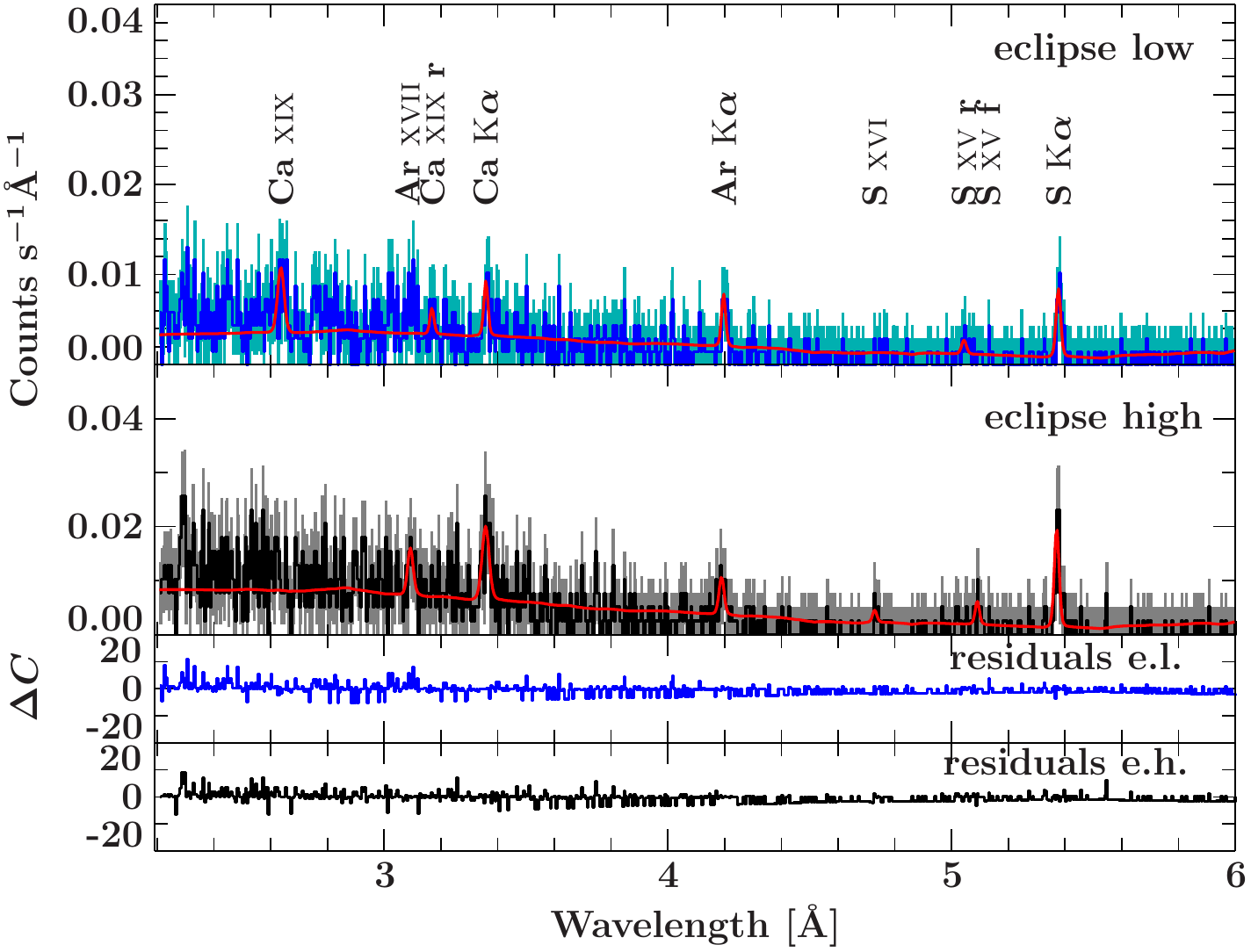}
\includegraphics[width=1\columnwidth]{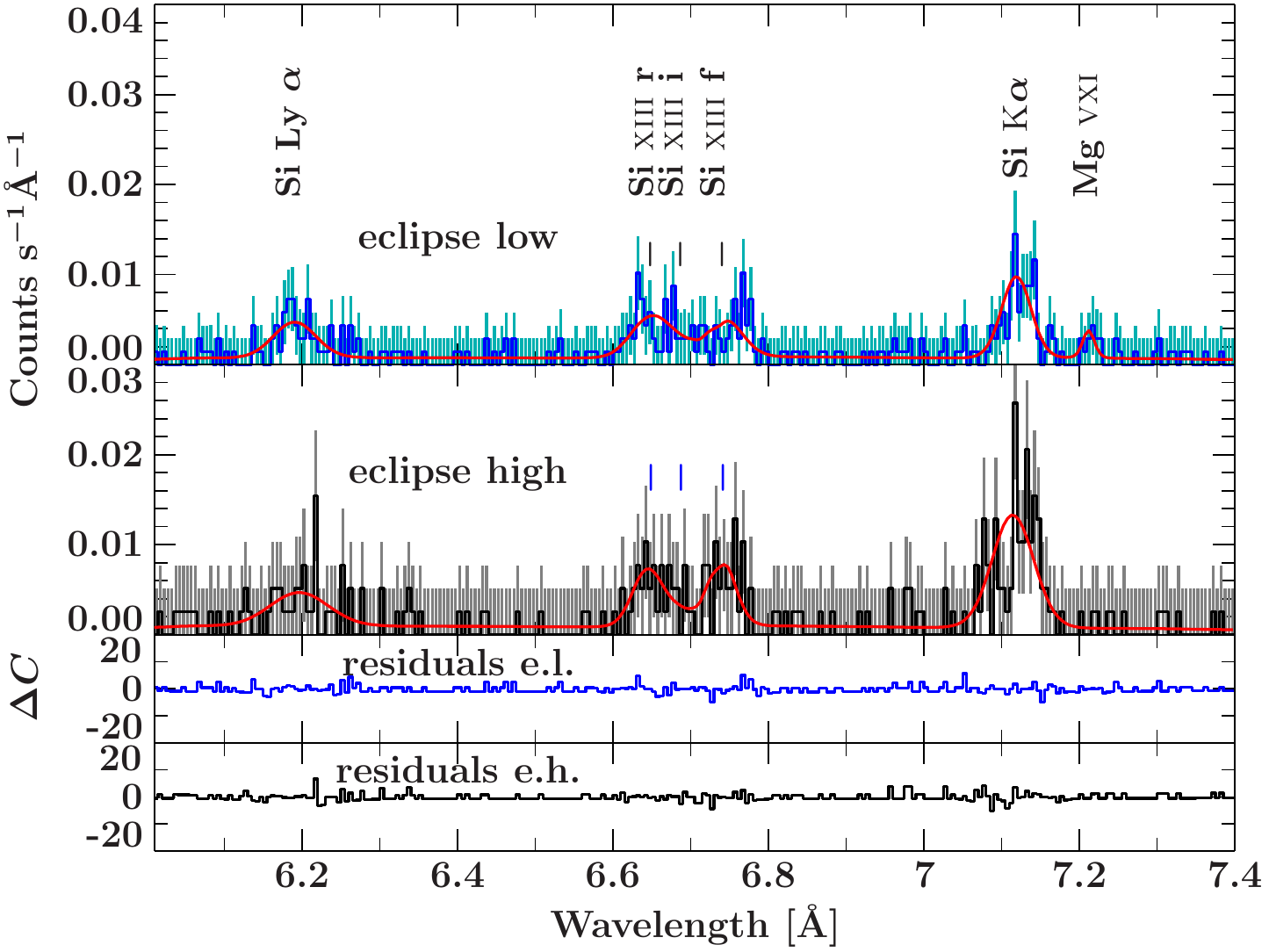}
\caption{\chandra Emission lines from 2.2$-$6 \AA\ (S) and 6$-$7.4 \AA\ (Si) regions. The model is shown in red. The error bars in light blue and the data in dark blue (eclipse low). The error bars in gray and the data in black (eclipse high).} 

\label{fig:S_LF_HF}
\end{figure*}

\begin{figure}
\includegraphics[width=0.99\columnwidth]{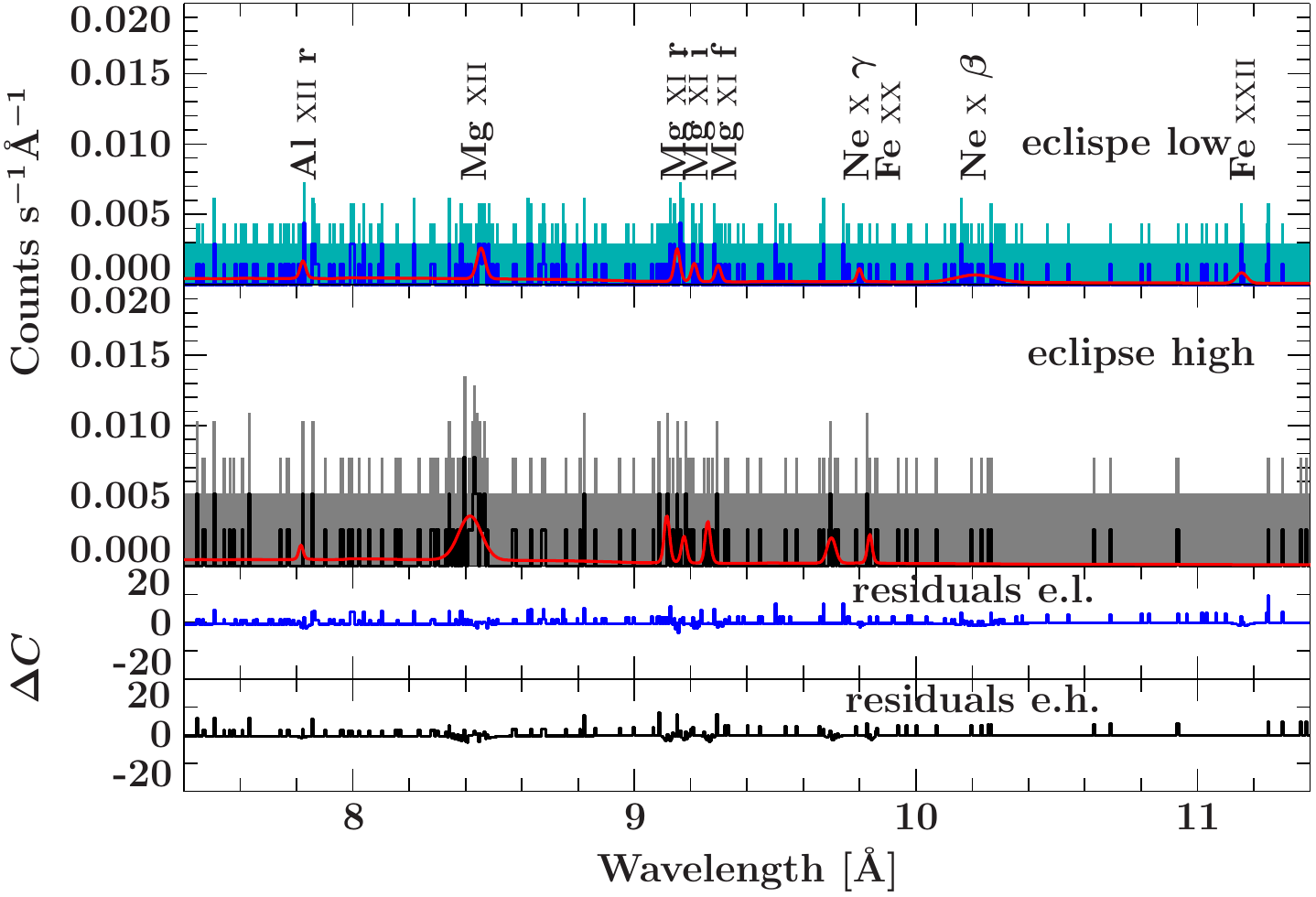}
\caption{\chandra Emission lines from 7.4$-$11.4 \AA\ (Mg) region.The model is shown in red. The errorbars in light blue an the data in dark blue (eclipse low). The errorbars in gray and the data in black (eclipse high).}
\label{fig:Mgregion}
\end{figure}

The line fit parameters are presented in Table~\ref{tab:Line_Emission_Eclipse}. Lines where $\alpha_{sig}$ is not quoted have $\alpha_{sig}<1.5$. Specially interesting sections are shown in Fig.~\ref{fig:Fe} (Fe line complex), Fig.~\ref{fig:S_LF_HF} (S and Si regions) and Fig.~\ref{fig:Mgregion} (Mg region).

On the other hand, in order to compare the line intensities during eclipse with their values out of eclipse, we will also use the spectral analysis from ObsID 17630. Details on the data extraction are given in \citet{2018MNRAS.473L..74M}. During the second half of the observation, the source flared, increasing its overall flux $\sim 6$ times. Consequently, the spectral extraction was divided into quiescence and flare. The line analysis has been carried out in the same way as described above. However, as the continuum here was significantly brighter, the significance of the Bayesian Blocks line detection was, generally, lower. The corresponding parameters are presented in Table \ref{tab:Line_Emission_Out_Eclipse}.

\subsection{Fluorescence Lines}

A number of K$\alpha$ fluorescence transitions from several elements (Fe, Ar, Ca, S and Si) are detected in the spectrum of \so (Fig. \ref{fig:Fe} and \ref{fig:S_LF_HF}). Fe~K$\alpha$ is the most prominent line in all our extractions for which the Bayesian Block algorithm obtains $\alpha_\mathrm{sig} > 100$ (Fig.~\ref{fig:Fe}). Fe~K$\beta$ is found by the Bayesian blocks algorithm only during eclipse high. However, we include and fit this line in eclipse low, as well.

All K$\alpha$ fluorescence line transitions respond directly to the continuum illumination so that their intensities grow with higher continuum fluxes (Table~\ref{tab:Line_Emission_Eclipse}). The line centroids, in turn, remain constant, within measurement uncertainties. For Fe~K$\alpha$,  
$\lambda=1.9404\pm0.0010$\,{\AA} in eclipse low and $1.9409\pm0.0009$\, \AA\ 
in eclipse high, compatible with \ion{Fe}{II-VIII}. The ratio between the line fluxes Fe K$\beta$ / Fe K$\alpha$ is $0.15^{+0.09}_{-0.07}$ in eclipse high and $0.11^{+0.08}_{-0.06}$ in eclipse low, consistent with the theoretical value of 0.13 for low optical depth lines \citep{1993A&AS...97..443K}.
The Fe K$\alpha$ line shows hints of asymmetrical profile in the high flux data. We attempt to describe the asymmetry by adding a second 
Gaussian component that could be interpreted as the Compton shoulder. However, the component is not significant, with an F-test resulting in 0.38\footnote{F-test must be used with caution when assessing the 
significance of an emission line \citep{2002ApJ...571..545P}}. Consequently, we do not include this component into further analysis and do not list it in Table \ref{tab:Line_Emission_Eclipse}. We point out that future high 
microcalorimeter resolution missions such as \textsl{XRISM} 
\citep{XRISM_2020a} and \textsl{Athena} \citep{Nandra_2013a} would allow a much more stringent test on the presence of a possible Compton Shoulder component.
The K$\alpha$ lines appear very narrow as their widths are not resolved at the \chandra \textsc{hetg} resolution (0.011 \AA). The exception is \ion{Si}{xiv}{} K$\alpha$, showing a $\sigma=0.018\pm 0.002$ \AA\ (eclipse low) and $\sigma=0.026\pm 0.005$ \AA\ (eclipse high). These correspond to plasma velocities $v=760$ km s$^{-1}$ and 1100 km s$^{-1}$ respectively. In fact, Silicon seems to be the only element whose width appears resolved, in all its detected ionisation states, as we will see in the next section.

\subsection{High-Ionisation Lines}
\label{sec:high_ion}
%%%TABLE G Y R CON  SILICON%%%%%%%%%%%%%%%%%
\begin{table}
\begin{center}
\caption{Parameters $G$ and $R$ 
%\MCHtext{MCH: I remove T(G) and Ne(R)}\redtext{VG: I think something went wrong with the table formatting or at least I don't know what you mean with Ne(R)x10 and T(G)x10$^{6}$K ... Sorry! }
\label{tab:GyRSi2}}
\begin{tabular}{cccc}
%\tableline\tableline
\hline

Ion & Parameter  & Eclipse High & Eclipse Low \\
%\tableline
\hline
 &$G$  & 1.3$^{+0.9}_{-0.7}$ &0.9$^{+0.6}_{-1.4}$ \\
 &$T_{\rm e}( \times10^{6}$ K) & 6.5$^{+3.5}_{-2.4}$&  7.5$^{+1.5}_{-1.5}$\\
  &$T_{\rm e}$ (keV) & 0.5$^{+0.3}_{-0.2}$&  0.6$^{+0.1}_{-0.1}$\\
Si &&&\\
 &$R$  & 3.4$^{+2.2}_{-1.7}$   &  2.9$^{+1.6}_{-1.4}$\\
 &$n_{\rm e}$ ($\times10^{13}$ cm$^{-3}$)& $<2$  & $<4$\\

%\tableline
\hline

\end{tabular}
\end{center}
\end{table}

A number of emission lines from highly ionised species are clearly seen in the eclipse spectrum. Of particular interest is the triplet consisting of
the transitions
$1\mathrm{s}^2\,^1\mathrm{S}_0\mathrm{-1s2p}\,^1\mathrm{P}_1$
(resonance, $r$),
$1\mathrm{s}^2\,^1\mathrm{S}_0\mathrm{-1s2p}\,^3\mathrm{P}_{2,1}$
(intercombination, $i$), and
$1\mathrm{s}^2\,^1\mathrm{S}_0\mathrm{-1s2s}\,^3\mathrm{S}_1$
(forbidden, $f$). Note, that the intercombination line splits up in
 two lines with upper levels $\mathrm{1s2p}\,^3\mathrm{P}_{1}$ and
$\mathrm{1s2p}\,^3\mathrm{P}_{2}$ but this splitting cannot be
resolved in the observation. The \ion{Fe}{xxv} $r$ and $f$ transitions are detected at low flux albeit with very low significance. No trace of \ion{Fe}{xxvi} is seen. This is in contrast with the observation out of eclipse (Table \ref{tab:Line_Emission_Out_Eclipse }) where both species are clearly detected during quiescence.

The most prominent lines are those of Si (Fig.\ref{fig:S_LF_HF} right) with the He-like \ion{Si}{xiii}{} triplet at $\lambda\sim 6.7$ \AA, being the strongest. Bayesian Blocks finds all three $r,i,f$ transitions, as one block with an $\alpha_{\mathrm{sig}}=23$ in eclipse high and $\alpha_{\mathrm{sig}}=30$ in eclipse low.

Within the uncertainties, the fluxes of the lines $r$ and $f$ are comparable. This is not expected in a purely photoionised plasma which would have $f > r$ and,  instead, suggests a low-density hybrid photoionised and/or collisionally ionised gas. \citet{2003ApJ...582..959W} suggest that photoionisation equilibrium still holds but that resonance scattering of the continuum by the $r$ lines adds to their flux during eclipse, when the direct continuum is suppressed. In any case, the forbidden transition $f$ is clearly detected, as seen in other HMXBs: Vela X-1 \citep{2002ApJ...564L..21S, 2017A&A...608A.143G}, Cygnus X-1 \citep{2019A&A...626A..64H}, 4U1538-52 \citep{2015ApJ...810..102T} and isolated O type stars \citep{2006ApJ...650.1096L, 2007ApJ...668..456W, 2015ApJ...809..132C}.

The presence of highly charged ions points to a very hot plasma. The kinetic energy of the continuum electron has to exceed the ionisation potential in order to be 
able to ionise the atom/ion in a collision. The ionisation potential 
to make H-like ions is fairly high. The electron 
temperature of the plasma has to be high enough to provide a sufficient 
number of electrons with kinetic energies above this threshold. 
Similarly, for collisional excitation of transitions, the kinetic energy 
of the colliding electron needs to exceed the excitation threshold energy.

Although the S/N is low (Fig.\ref{fig:S_LF_HF}), we can obtain  the parameters $G= (i + f )/r$ and $R= f/i$ for the \ion{Si}{XIII} triplet which provide direct estimations of plasma temperature and density \citep{2000RMxAC...9..316P}. In O stars, though, the strong photospheric UV continuum depopulates $f$ into $i$ thereby changing the ratios above and the corresponding deduced plasma properties (see Section \ref{sec:disc_highion}). We obtained $R$ and $G$ implementing the formulas above as functions directly into the fitting programme, so that we could also directly obtain the errors. The corresponding parameters are presented in Table~\ref{tab:GyRSi2}. The plasma temperatures are of the order of several million degrees.

\subsection{Photo ionised plasma models}

%%%%%%%%%%%%% PLOT PHOTEMIS%%%%%%%%%%%%%%%%%%%%%%%%
%%%%%%%%%%%%%%%%%%%%%%%%%%%%%%%%%%%%%%%%%%%%%%%
\begin{figure*}

\includegraphics[width=2.07\columnwidth]{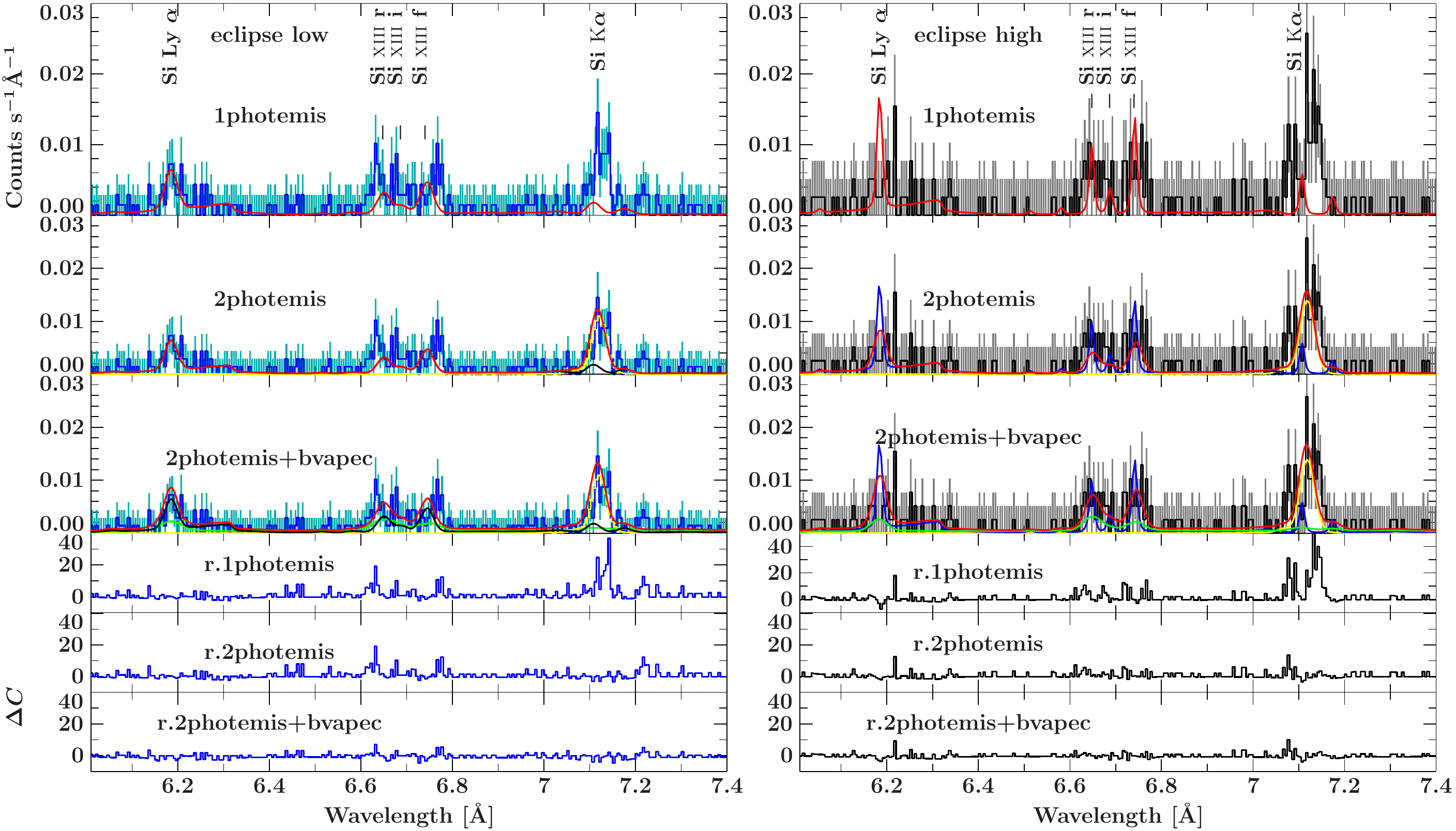}
\caption{\chandra emission lines in the 6$-$7.4 \AA\ range for low (left, blue) and high flux (right, black) periods during the eclipse. The first and second panels shows the fit using only one or two photoionised plasma (\texttt{photemis}) components, respectively. The model is shown in red. Second panel: (Eclispe low) contribution to the model from \texttt{photemis} with $\log \xi\sim2.38$ (black) and with $\log \xi\sim -1.0$ (yellow); (Eclispe high) contribution to the model from \texttt{photemis} with $\log \xi\sim2.34$ (blue) and with $\log \xi\sim-1.04$ (yellow). Third panel shows the best fit for the above two \texttt{photemis} plus a collisionally ionised plasma \texttt{bvapec} (green).} 
\label{fig:photemis}
\end{figure*}

Apart from the above phenomenological approach, we also tried to model the spectrum with a self consistent photoionised plasma emission model. We focus on the Silicon region for this study, from 6.0 to 7.4 (\AA) because it has the highest significance among the triplets. To that end, we use \texttt{photemis} based on \textsc{xstar}\footnote{https://space.mit.edu/cxc/analysis/xstardb/index.html} \citep{Bautista_2001a, Kallman_2001a}. \texttt{photemis} is the ''thermal'' (i.e. recombination and collisional excitation) emission which comes from the analytic plasma model that allows the use of warm absorbers and photoionised emitters, as well as for coronal equilibrium absorbers and emitters and  employs the most recent updates to \textsc{xstar}. We further add a simple powerlaw to model the local continuum. We note that \texttt{photemis} models are calculated for a power law illumination with $\Gamma = 2$ and do not include the influence of the strong UV emission from the star, that may change the contribution between the components of the triplet (see Sec.~\ref{sec:disc_highion}). We have shown the influence of UV emission in the HMXB Vela X-1  \citep{Lomaeva_2020a}. 

As we have seen in the previous sections, highly ionised species coexist with low ionisation or near neutral ones. Thus, logically, a single \texttt{photemis} can not satisfactorily reproduce the whole spectrum (Fig. \ref{fig:photemis}, upper panel), yielding $C^2_r$ of 2.43 and 2.01 for eclipse low and high, respectively.  Two plasmas are needed, one with low ionisation ($\log\xi\sim -1)$ and other with high ionisation ($\log\xi\sim 2.4)$, where $\log\xi$ is the ionisation parameter (see Discussion). However, although they fit well the spectrum overall, some line profiles can not be reproduced. This is clearly seen in Fig. \ref{fig:photemis} (middle panel). The \ion{Si}{xiii}{} triplet shows a particularly complicated profile. It appears to be formed by four narrow lines none of which is neither at the lab rest frame $\lambda$ (marked by vertical lines below the $rif$ label transitions) nor are they shifted, all together, in a particular direction. The fit then requires broadening the lines with a bulk plasma velocity $v_{\mathrm{turb}}\sim 840$\,km\,s$^{-1}$.  In any case, the photoionisation models predict $f>r$ whereas the data shows $f\approx r$ (Fig. \ref{fig:S_LF_HF}, right panel and Table \ref{tab:Line_Emission_Eclipse}). 

Adding a third photoionised plasma does not help. In fact, the only way of approaching the observed ratio is by adding a third \textsl{collisionally} ionised plasma. For this purpose we use  \texttt{bvapec}, a velocity and thermally-broadened emission spectrum from collisionally-ionised diffuse gas calculated from the AtomDB atomic database. Its temperature turns out to be $kT \approx 1$\,keV (Table \ref{tab:photemis}). Although the resulting statistic is now acceptable ($C^2_r$ equal to 1.11 and 1.04 for eclipse low and high respectively), significant residuals still remain, particularly during the low state, thus demonstrating the complexity of the \ion{Si}{xiii} triplet and, possibly, the multiple origin of plasmas contributing to the observed spectrum.

\begin{table}
{\def\arraystretch{1.3}
\begin{center}
\caption{\texttt{photemis} model best parameters.}
\begin{threeparttable}
\label{tab:photemis}
\begin{tabular}{lcc}
\hline
Parameter &  Eclipse Low  & Eclipse High \\
\hline

&\texttt{photemis 1}&\\

Norm 		&	$31\pm2$	&	$40\pm1$		\\
$\log\xi$ 		&	$2.38\pm0.01$	&	$2.34\pm0.01$		\\
$v_{\rm turb}$ (km\,s$^{-1}$)        &	$830\pm2$ &  $845\pm1$	\\

&\texttt{photemis 2}&\\

Norm (x10$^{5}$)	&	$1.6^{+0.4}_{-0.3}$	&	$2.3\pm1$		\\

$\log\xi$ 		&	$-1.0\pm0.1$	&	$-1.04\pm0.04$\\
$v_{\rm turb}$ (km\,s$^{-1}$)        &	$830\pm2$ &  $845\pm1$	\\
&&\\							
 C$^{2}_{\rm r}$(d.o.f.)		&	1.44(234)	&1.18	(234)		\\
\hline\hline
&\texttt{bvapec} &\\ 
\hline
Norm $(\times 10^{-4})$ & $10^{+1}_{-3}$ & $9^{+4}_{-2}$\\
%$kT$ (keV) & $1.02\pm0.19$ &  $1.0^{+0.2}_{-0.3}$\\
 $T_{\rm e}\times10^{6}$ K & 11$^{+2}_{-3}$&  $11\pm2$\\
$v$ $(\times10^{2}$\,km\,s$^{-1})$ &  $11\pm3$ &  $8\pm3$\\
&&\\
C$^{2}_{\rm r}$(d.o.f.)		&	1.11(217)	&1.04	(217)		\\
\hline\hline

	\end{tabular}
\begin{tablenotes}
\item[a] Low flux: $F$-test$=$9.0e-07; High Flux: $F$-test$=$0.04.

\end{tablenotes}
\end{threeparttable}
\end{center}
}
\end{table}

\section{Discussion}
\label{sec:disc}

\begin{figure*}
\includegraphics[width=2\columnwidth]{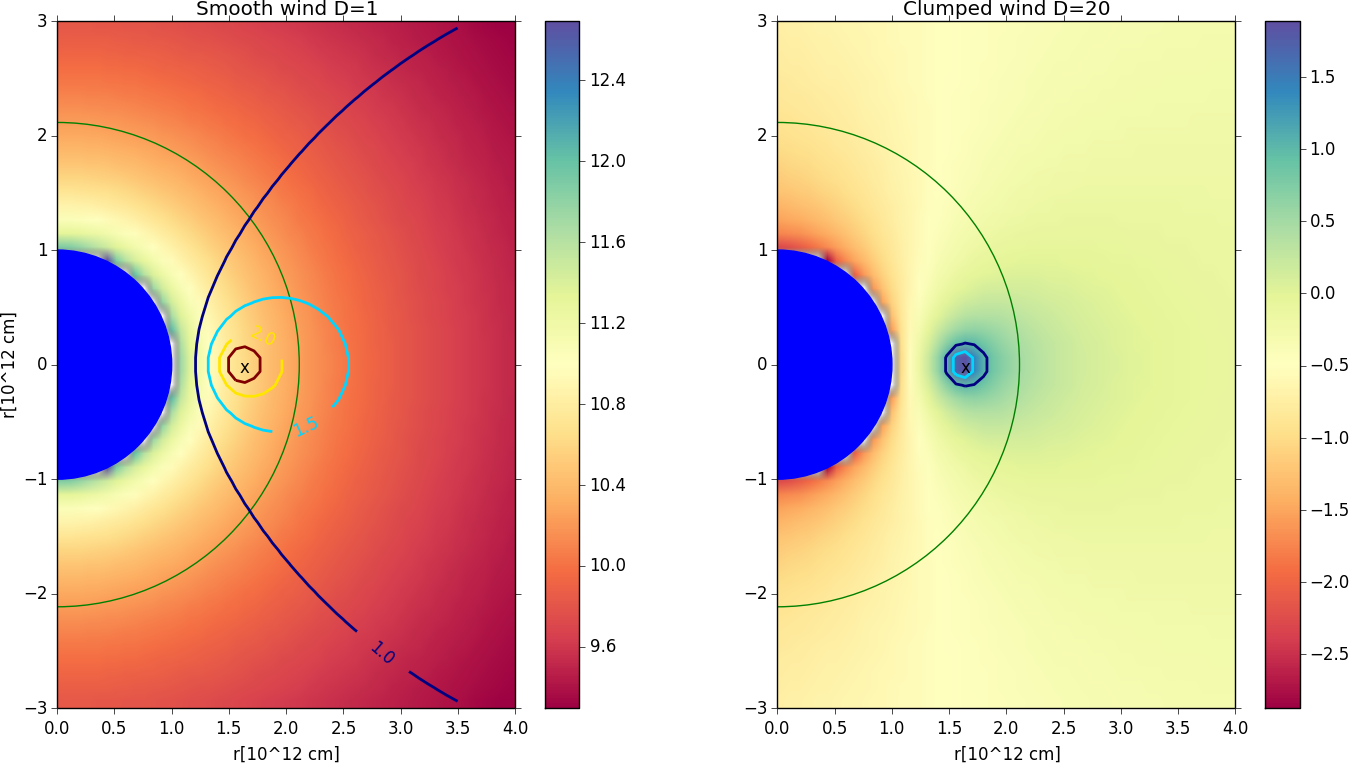}
\caption{Scheme of the system. The blue circle is the donor star while the X at three o'clock marks the position of the compact object. The green circle marks the radial distance at which the wind reaches 800 km s$^{-1}$. For reference, background shows $\log n$ particle density [cm$^{-3}$] (left) and $\log \xi$ (right), for a smooth wind ($D=1$).  Contours of iso-ionisation are presented for a smooth ($D=1$) and a clumped wind ($D=20$), respectively.}
\label{fig:n_logxi}
\end{figure*}

\subsection{Light curve variability}

As we have seen in Section \ref{sec:lc}, the optical-NIR and the X-ray lightcurves show significant variations throughout the observation. On one hand, the X-ray lightcurve displays a series of flares at the beginning of the eclipse, when the neutron star is completely hidden from view. In fact, such flares can be observed over the entire eclipse duration \citep[][Fig. 2]{2016MNRAS.461..816I}. The flares must be produced at the X-ray source (e.g. the hot spots and the accretion columns on the neutron star surface). Since these are blocked from direct view, the light must be reflected into the observers line of sight by some dense structure, comparable in size with the donor's radius $\sim 10^{12}$ cm $\sim R_{*}$ or the donors wind. This structure could be also reflecting star light: once it is eclipsed, the optical magnitude reaches a constant value, corresponding to the star brightness. Such an enormous structure would probably be detected in the optical-UV spectrum. \citet{1994A&A...289..846K} found strong evidence that the ionisation wake in \so is causing orbital modulated absorption in its optical lines. Also, large scale structures in O star winds (corotating interaction regions or CIRs) are commonly detected in UV, optical and X-rays \citep{2008ApJ...678..408L, 2014MNRAS.441..910R, 2019ApJ...873...81M}. 

Another possibility is that the X-ray reflecting structure is the bulk of the donor's wind. In such a case, the extra optical light could come from the reprocessing of X-rays in the donor's photosphere. Such a {\it hot spot} would also be detected in the optical spectrum at the right orbital phases. However, \citet{1978Natur.275..400D} did not find any significant difference between the \textit{IUE} spectra of \so at orbital phases $\phi =0.32$ and $\phi =0.97$ (neutron star eclipse). Likewise, our \textit{HST} observations \citep{2020A&A...634A..49H}, obtained simultaneously at the end of ObsID17630 ($\phi\approx 0.16$) find stellar and wind parameters normal for the spectral type. Dedicated observations will be needed to verify these hypotheses.

\subsection{Emission line spectra}

\subsubsection{Fluorescent Lines}
The K shell fluorescence lines from (near) neutral species directly correlate with the X-ray source brightness. These transitions can arise in the dense cold clumps of the stellar wind \citep[][]{1999ApJ...525..921S} and/or in accretion stream structures \citep[i.e.][]{2019ApJS..243...29A}. Their intensities are compatible, within the uncertainties, for the eclipse (ObsID~18951) and out of eclipse (ObsID~17630) observations. For example, the intensity of the Fe~K$\alpha$ line is $(330^{+120}_{-90})\times 10^{-6}$\,ph\,s$^{-1}$\,cm$^{-2}$ out of eclipse (quiescence, Table~\ref{tab:Line_Emission_Out_Eclipse }) and $(270\pm40)\times 10^{-6}$\,ph\,s$^{-1}$\,cm$^{-2}$ during eclipse low (Table~\ref{tab:Line_Emission_Eclipse}). The same is true for S and Si K$\alpha$. However, if we compare our eclipse low value with the out-of-eclipse ObsID 657, namely, $(870\pm 81)\times 10^{-6}$\,ph\,s$^{-1}$\,cm$^{-2}$ \citep[][]{2010ApJ...715..947T}, the ratio turns out to be $\sim 0.3$, in line with those obtained using \textit{XMM-Newton} \citep[][]{2015A&A...576A.108G, 2019ApJS..243...29A} and also with other HMXBs with supergiant donors with later spectral types. For example, in QV Nor (B0.5Ib), the donor of the X-ray pulsar 4U1538$-$52, the Fe K$\alpha$ line is reduced to 0.3 of its original intensity during an eclipse \citep{2015ApJ...810..102T}\footnote{The continuum, however, is reduced to a 0.1 its original intensity which, by contrast, increases the equivalent width of the Fe line.}. A similar behavior is seen in Vela X-1 \citep{Goldstein_2004a,2010ApJ...715..947T}. Thus it is intriguing that comparing source states, within each observation, that should be, in principle, similar, in the sense that they are out of flaring (quiescence and eclipse low), the Fe K$\alpha$ lines show similar intensities. In fact, \cite{2003ApJ...592..516B}, also from ObsID 657, quote an intensity for Fe K$\alpha$, during quiescence, of $(170\pm 60)\times 10^{-6}$\,ph\,s$^{-1}$\,cm$^{-2}$ lower than, but compatible, with our measurements. 

Photons emitted in fluorescence lines can not be resonantly scattered in the wind because they do not have enough energy to induce further extractions from the inner K shell of the atoms. Thus, all fluorescence emission must be produced at sites directly in the line of sight of the observer (and the compact object). This means that, while in QV Nor the vast majority of Fe K$\alpha$ photons were produced close to the donor's photosphere, facing the neutron star, in \so they have to be produced mostly at radial distances $r_{\rm X}> R_{*}$ from the compact object so that their intensity does not decrease much during eclipse. 

In Fig. \ref{fig:n_logxi} we map the ionisation parameter $\log \xi$, where
\begin{equation}
\xi = \frac{L_{x}}{n(r_{x})r^{2}_{x}}
\label{eq:ionization_parameter}
\end{equation}
with $n(r_{x})$ the number density of atoms at radial distance $r_{X}$ from the X-ray source and $L_{x}$ the X-ray luminosity of the source. To account for wind clumping we use the density contrast $D=\rho_{\rm cl}/\Bar{\rho}$, where $\rho_{\rm cl}$ is  the  density  of  the clumped medium and $\Bar{\rho}$ the average density. The inter-clump medium is  assumed  to  be  empty.  $D=1$ corresponds to smooth plasma.

We also show some iso-ionisation contours. For the calculation of $\log \xi$, we have used the parameters in \citet{2020A&A...634A..49H} (Table~\ref{tab:Parameters system 2}) and a radial density distribution based on the mass continuity equation $\rho(r) = \dot{M}/4\pi v(r) r^2$ with $v(r)$ following a \textsl{double} beta velocity law\footnote{$v(r)=v_\infty\left[0.6(1-\frac{R_{*}}{r})^{0.8}+0.4(1-\frac{R_{*}}{r})^{\beta}\right]$} (see \citealt{2020A&A...634A..49H} for details). The ionisation parameter given by Eq. \ref{eq:ionization_parameter} is reduced by a factor $D$ for the clumped wind case \citep{2012MNRAS.421.2820O}. For the source luminosity $L_{\rm X}$ we have used the out of eclipse quiescence value obtained in \citet{2018MNRAS.473L..74M}, namely, $L_{\rm X}\approx 3\times 10^{35}$ erg s$^{-1}$\footnote{for $d=1.7$ kpc,   $L_{\rm X}\approx$ $[0.1-8]\times 10^{36}$ erg s$^{-1}$ \citet{1989ApJ...343..409H},  $\approx$ $0.2\times10^{36}$ \citet{2003ApJ...592..516B}, $\approx3.6\times10^{35}$ \citet{2005A&A...432..999V}}. In both cases, the ionisation parameter is rather low ($<2$) for most of the wind. Neutral species can exist, and therefore K$\alpha$ fluorescence transitions be excited by X-ray photons from the compact object, in the whole wind. The wind velocity compatible with the narrow width of Fe K$\alpha$ is attained at $r\leqslant 2.12R_{*}$. The wind in \so (O6.5Ia) is thicker than in QV Nor (B0.5Ib), with a density radially decreasing at much lower rate (since here $\beta = 2\pm 1$). Thus, it is expected that transitions corresponding to neutral species will not decrease much during eclipse.

\subsubsection{High-Ionisation Lines}
\label{sec:disc_highion}

The analysis of optical-UV spectra shows that the stellar wind of \so must be clumped \citep{2020A&A...634A..49H}. As discussed in Section~\ref{sec:analysis}, the K$\alpha$ fluorescence transitions, produced in the dense cold parts of the stellar, must be coexisting with a highly ionised plasma. While the former is associated with the overdense structures in the stellar wind, a.k.a. wind clumps, the latter must arise, presumably, in the rarefied and hot interclump medium \citep[][for the case of Vela X-1]{1999ApJ...525..921S}. The fits to photoionisation plasma models allow us to gain insight into the density contrast between both. The normalization of the \texttt{photemis} model is $N_{\texttt{phot}}=10^{-10}EM/4\pi d^{2}$ where $EM$ is the plasma emission measure. Using values in Table \ref{tab:photemis}, $EM_1=EM_{\rm hot}=1.2 \times 10^{56}$ cm$^{-3}$ (we associate the hot plasma with the interclump medium) while $EM_2=EM_{\rm cold}=5.5 \times 10^{59}$ cm$^{-3}$. Thus, the ratio $N_2/N_1\sim EM_{\rm clumps}/EM_{\rm interclump} = EM_{\rm c}/EM_{\rm i}\approx 5\times 10^{3}$. Now, $EM \sim n^2 V$, where $n$ and $V$ are the density and 
volume of the emitting plasma. In a simplified case of a two phase media consisting of the 
clumps and interclump gas (each of constant density), $EM_{\rm c} \sim n_{\rm c}^2 V_{\rm c}$ and $EM_{\rm i} \sim n_{\rm i}^2 V_{\rm i}$, and 
\begin{equation}
 \frac{EM_{\rm c}}{EM_{\rm i}}=\frac{n_{\rm c}^2}{n_{\rm i}^2}\frac{V_{\rm c}}{V_{\rm i}}.
 \label{eq:emv}
\end{equation}

\citet{2020A&A...634A..49H} determined the clumping factor $D$ which 
describes by how much density in clumps is 
enhanced compared to the density of a smooth wind, $n_{\rm w}$, with the same mass-loss rate. That is to say, $n_{\rm c} = D n_{\rm w}$. Factor $D$ is derived from the fitting of UV and optical spectra 
(Table\,1) assuming that the interclump medium is void (thus $n_{\rm i} = 0$). This is suitable for the analysis since we assume that the interclump medium does not contribute to the emission in UV and optical. By analogy, lets define the parameter $d$ describing by how much density is reduced in the interclump 
medium, $n_{\rm i} = d n_{\rm w}$. Recalling that $V_{\rm wind}=V_{\rm c}+V_{\rm i}$, 
the clump volume filling factor becomes 
\citep[see Eq.\,(19) in ][]{2012A&A...541A..37S}
\begin{equation}
 f_{\rm V}\equiv \frac{V_{\rm c}}{V_{\rm wind}}=\frac{1-d}{D-d}.
 \label{eq:fv}
\end{equation}
Then, the emission measure ratio determined from the observations
can be expressed as 
\begin{equation}
 \frac{EM_{\rm c}}{EM_{\rm i}}=\frac{D^2}{d^2}\frac{f_{\rm V}}{1-f_{\rm V}}.
 \label{eq:emr}
\end{equation}
Combining Eq.\,(\ref{eq:fv}) and (\ref{eq:emr}), 
\begin{equation}
 \frac{EM_{\rm c}}{EM_{\rm i}}= \frac{D^2}{D-1}\frac{1-d}{d^2}.
\end{equation}
Since emission measure ratio and the clumping factor $D\sim 20$ are known, we can
solve for $d$ yielding $d\approx0.06$, and the density ratio between clumps and interclump medium, $n_c/n_i=D/d\approx 330$. The clumps volume filling factor $f_{\rm V}$ is then $\approx 0.05$ (Eq. \ref{eq:fv}), very similar to that found for Vela X-1 by \cite[][$f_{\rm V}\approx 0.04$]{1999ApJ...525..921S}. In summary, while the wind mass is dominated by the clumps, 95\% of the volume is occupied by the hot interclump medium, with a density contrast between them of several hundred.

%Now $EM\sim n^{2}V$, where $V$ is the emitting plasma volume. The wind volume occupied by clumps and interclump medium must be essentially the same at large scale. Then, the ratio $EM_{\rm clumps}/EM_{\rm interclump}\sim (n_c/n_i)^{2}$. Thus, the clumps-to-interclump medium density ratio must be $n_c/n_i\sim 70$. This is largely compatible with the upper limit on clumping factor as assumed by \citet{2020A&A...634A..49H} to fit the optical/UV spectra.

On the other hand, the normalization of the collisionally ionised component, \texttt{bvapec}, is $\sim 9\times 10^{-4}$. In this case $N_{\texttt{bvapec}}=(EM/4\pi d^{2})10^{-14}$. Therefore, $EM_{\texttt{bvapec}}\approx 3.11\times 10^{55}$ cm$^{-3}$. For typical wind densities of the order of $n\sim 10^{10-11}$ cm$^{-3}$, the plasma would have a characteristic size $r\sim 10^{10-11}$ cm $=[\frac{1}{100},\frac{1}{10}]R_{*}$. In other words, the collisionally ionised component is very localized within the system. \so thus displays a hybrid 
plasma with photo-ionised and collisionally ionised contributions. 

In a plasma strongly influenced by UV radiation, as expected from the hot photosphere of the star, the $f/i$ ratio decreases due to the depopulation of $f$ line into $i$ \citep{1969MNRAS.145..241G}. Thus, the $R=f/i$ parameter tends to decrease, mimicking a higher density plasma. The \ion{Si}{xiii} triplet, by far the one showing the strongest signal in the eclipse spectrum, clearly shows $f>i$ regardless of continuum illumination. This behavior is less clear for \ion{Mg}{xi} due to the much higher uncertainties. Thus, the density values quoted in Table \ref{tab:GyRSi2} ($n_{\rm e}\sim 3\times 10^{13}$ cm$^{-3}$) must be taken as upper limits. Additionally, the analysis of the \textsl{HST-UV} spectra \citep{2020A&A...634A..49H}, during which an X-ray flare occurred \citep{2018MNRAS.473L..74M}, found stellar and wind parameters consistent with the ones expected for the donor's spectral type. The stellar wind thus does not seem to be modified or perturbed at large scale by the neutron star's X-ray emission.

Finally, highly ionised Fe\textsc{xxv} and Fe\textsc{xxvi} requires $\log \xi\sim 3$, a condition which is only met very close to the neutron star (Fig. \ref{fig:n_logxi}). Therefore these lines will be greatly diminished during eclipse, as observed\footnote{Some photons will still be visible during eclipse due to resonant scattering in the wind.}. They will also be more prone to react to changes in the X-ray continuum \citep{2018MNRAS.473L..74M}.

We can estimate also the maximum radius of line formation. We have measured, or set upper limits to, the width $\sigma$ of the emission lines. Assuming that the line is broadened by the bulk motion of the stellar wind, we can calculate the corresponding wind velocity $v_{\rm w}=c\sigma/\lambda $ for each ion. Then, using the parameters in Table \ref{tab:Parameters system 2} ($v_{\infty}=1900$ km s$^{-1}$ and $\beta=2$) we can estimate the maximum formation radius $r_{\rm max}$, assuming a wind velocity profile type double {\it beta law}, as described above. These values are presented in Table \ref{tab:velocities}. 

%%%%%TABLE WITH ION PROPERTIES MNRAS %%%%%%%%%%%%%%%%%%
\begin{table}
{\def\arraystretch{1.3}
\begin{center}
\caption{Radial range formation radius of the different ions. The values correspond to the eclipse low state. For the triplets, only the $r$ transition is quoted.}
\begin{threeparttable}
\label{tab:velocities}

\begin{tabular}{lccc}
\hline
Ion  &    $v_{\rm w}$             & $r_{\rm min}$ & $r_{\rm max}$  \\
      &  (km\,s$^{-1}$)  & ($R_{*}$) & ($R_{*}$) \\
\hline
%& & & & & &\\
%\ion{Fe}{ \ K$\beta$} 	&	  		&		&	\\		
																										
%\ion{Fe}{xxv $r$}	&	  		&	 &\\										
																										
\ion{Fe}{ \ K$\alpha$} 	&	   770		&	1.6	&	4.1\\		
																										
%\ion{Ca}{xix}   	&	  		&		&	\\	
																										
%\ion{Ar}{xvii}   	& &1.2	& 2.4\\	

\ion{Ca}{xix$r$}	&	 1140		&	2.3	& 7.1	\\	
																										
\ion{Ca}{\ K$\alpha$}  	&	  450		&	1.3 &	2.7	\\	
																										
\ion{Ar}{\ K$\alpha$} 	&	   360		&	1.2 &	2.4\\	
																										
\ion{S}{xv $r$}	&	   360		&		1.2 & 2.4	\\									

\ion{S}{\ K$\alpha$ }	&	  280		&	1.2 & 2.2	\\	
																										
\ion{Si}{xiv Ly$\alpha$} 	&	   1260		&	2.7	& 8.9	\\	
																										
\ion{Si}{xiii $r$}	&	  590		&	1.4	&	3.2\\	
																										
\ion{Fe}{xxiv} 	&	   360	&	1.2 &	2.4	\\	
																										
\ion{Si}{xiv K$\alpha$ }	&	   930	&		1.9 &	5\\	
																										
\ion{Mg}{xi} 	&	   210		&	1.1	&	1.9	\\	

\hline
\end{tabular}																		
																		
\end{threeparttable}	

\end{center}																		
}																		
\end{table}

%%%%%%%%%%%%END TABLE %%%%%%%%%%%%%%%

The majority of line widths can not be resolved even at the \chandra-\textsc{hetg} resolution, locating the formation region relatively close to the donor's photosphere. The exception is \ion{Si}{xiv}{} Ly$\alpha$ which extends up to 4$R_{*}$ while it can reach up to $r_{\rm max}=24R_{*}$ during high flux. There is no systematic separation between the neutral species and their highly ionised counterparts. Cold and hot wind phases must coexist at the same radial distances within the wind, consistent with the view of cold dense clumps interspersed in a hot rarefied interclump medium.

\section{Conclusions}
\label{sec:conc}
We have presented an analysis of the first observation of \so with \chandra High Energy Transmission Gratings during eclipse. This allow us to study in depth the back illuminated stellar wind structure and properties of the O6Ia star HD153919, the earliest donor in any HMXB, with unprecedented detail. We find that: 
\begin{enumerate}
\item emission lines from K shell transitions, corresponding to near neutral species, increase their brightness in response to an increased continuum illumination. However, they do not greatly diminish during eclipse, in contrast with other HMXBs with later type donors. This is readily explained if fluorescence K$\alpha$ emission comes from the bulk of the wind. 

\item In contrast, the highly ionised \ion{Fe}{xxv}{} He-like and \ion{Fe}{xxvi}{} Ly$\alpha$ must be produced in the vicinity of the compact object, the only region where the ionisation parameter is sufficiently high, $\log \xi >3$. Therefore, these lines diminish greatly during eclipse (\ion{Fe}{xxvi}{ Ly$\alpha$} is not detected in eclipse).

\item the addition of two self consistent photo ionisation models \texttt{photemis}, from \textsc{xstar}, with low ionisation ($\log \xi\sim -1$) and high ionisation ($\log \xi\sim 2.4$) degrees respectively, are required to describe the emission line spectrum. From their emission measures, and the clumping factor deduced from the optical-UV spectra, the clump-to-interclump density ratio can be estimated to be $n_c/n_i\sim 330$. However they are not able to fit the shape of the He-like \ion{Si}{xiii}{} which shows a complex structure. Statistically, the fit requires line broadening with $v_{\rm bulk}\sim 840$ km/s. Furthermore, to reproduce the observed $r\approx f$ fluxes, the addition of a collisionally ionised plasma, with $kT\sim 1$ keV, is required. The emission measure of this component, however, points to a rather small plasma volume.

\item All detected emission lines widths appear unresolved at the \chandra \textsc{hetg} gratings resolution. The exception is Silicon. Assuming that the main broadening mechanism is the bulk plasma velocity, \ion{Si}~K$\alpha$ shows a range 800--1000\,km\,s$^{-1}$. On the other hand, \ion{Si}{xiv}~Ly$\alpha$ shows a range 1300--1800\,km\,s$^{-1}$. There is no clear radial segregation between (quasi)neutral and ionised species. This is consistent with the picture of cold wind clumps interspersed in a hot rarefied interclump medium.

\end{enumerate}

\section*{Acknowledgements}

This research has been funded under the project ESP2017-85691-P. The research leading to these results has received funding from the European Union's Horizon 2020 Programme under the AHEAD project (grant agreement n. 654215). VG is supported through the Margarete von Wrangell fellowship
  by the ESF and the Ministry of Science, Research and the Arts
  Baden-W\"urttemberg. This research has
  made use of ISIS functions (\texttt{isisscripts})\footnote{\url{http://www.sternwarte.uni-erlangen.de/isis/}}
  provided by ECAP/Remeis observatory and MIT and of NASA's
  Astrophysics Data System Bibliographic Service (ADS). This research has made use of ISIS functions (\texttt{xstardb}) provided by the MIT Kavli Institute for Astrophysics and Space Research  (\url{http:/space.mit.edu/cxc/analysis/xstardb})." We thank John
  E.  Davis for the development of the
  \texttt{slxfig}\footnote{\url{http://www.jedsoft.org/fun/slxfig/}}
  module used to prepare most of the figures in this work.
  %NATALIE:
  Work at LLNL was performed under the auspices of the U.S. Department of 
Energy under contract No. DE-AC52-07NA27344 and supported through NASA grants to LLNL. We thank the anonymous referee whose comments improved the content of the paper. 

%%%%%%%%%%%%%%%%%%%%%%%%%%%%%%%%%%%%%%%%%%%%%%%%%%
\section*{Data Availability}

 The data used in this paper is publicly available at the \chandra archive, https://cda.harvard.edu/, with the identifiers ObsID 17630 and ObsID 18951.



%\bibliographystyle{mnras}

\bibliographystyle{mnras}
\bibliography{bibliografia}

%\end{document}

%%%%%%%%%%%%%%%%% APPENDICES %%%%%%%%%%%%%%%%%%%%%
%\newpage
\appendix
%%%%%%%%%%%%%%%%%%%%%%%%%%%%%%%%%%%%%%%%%%%%%%%%%%
\newpage
\section{Emission line parameters}

\begin{table*}
{\def\arraystretch{1.4}
%\begin{center}
\caption{Eclipse emission lines (ObsID 18951).\label{Line Emission }}
\begin{threeparttable}
\label{tab:Line_Emission_Eclipse}

\begin{tabular}{lcccccccc}
\hline
& &Eclipse Low & & &  & Eclipse High & &\\

\hline

Line  &    $\lambda$             &  Line Flux        & $\sigma$   &Bayesian   & $\lambda$ &  Line Flux          & $\sigma$&Bayesian  \\
      &                          &  $\times 10^{-6}$  &          & Blocks     &           &    $\times 10^{-6}$ &         & Blocks   \\
      & (\AA)                    & (ph s$^{-1}$ cm$^{-2}$)  &(\AA) & $\alpha$$_\mathrm{sig}$     &  (\AA)&  (ph s$^{-1}$ cm$^{-2}$)& (\AA)      &  $\alpha$$_\mathrm{sig}$ \\
\hline													
\ion{Fe}{\ K$\beta$}	&	$1.760^{+0.008}_{-0.005}$		&		  $31^{+22}_{-17}$		&	0.005	$^{+0.003}_{-0.000}$		&		&	$1.761^{+0.007}_{-0.006}$		&	   $100^{+60}_{-50}$		&	$0.008^{+0.002}_{-0.003}$			&	3		\\	
																												
 \ion{Fe}{xxv $r$}	&	$1.848^{+0.012}_{-0.019}$		&		  $13^{+18}_{-13}$		&	0.005	$^{+0.003}_{-0.000}$				&													\\	
  \ion{Fe}{xxv $f$}	&	$1.865^{+0.012}_{-0.019}$		&		  $8^{+14}_{-9}$		&	0.005	$^{+0.003}_{-0.000}$		&	2	&	$1.8674^{+0.0027}_{-0.0075}$		&	   $33^{+30}_{-24}$		&	$0.008^{+0.002}_{-0.003}$			&			\\	
																										
\ion{Fe}{ \ K$\alpha$} (single)	&	$1.9404\pm0.0010$		&		   $270\pm40$		&	0.005	$^{+0.001}_{-0.001}$		&	100	&	$1.9409\pm0.0009$		&	   $680\pm70$		&	$0.006^{+0.001}_{-0.001}$			&	100		\\	
																												
\ion{Ca}{xix}   	&	$2.636^{+0.008}_{-0.009}$		&		  $15^{+7}_{-6}$		&	0.01			&	2.2	&			&			&				&			\\

\ion{Ar}{xvii}   												&			&&&&$3.092\pm0.009$		&	   $12^{+9}_{-8}$		&	0.008	$^{+0.001}_{-0.003}$		&			\\	
\ion{Ca}{xix}$r$	&	$3.168^{+0.011}_{-0.009}$		&		 $3.2^{+3.7}_{-2.9}$		&	0.005			&	2.2	&			&			&				&			\\

\ion{Ca}{\ K$\alpha$}   	&	$3.358^{+0.009}_{-0.006}$		&		   $6^{+5}_{-4}$		&	0.005			&	-	&	$3.357\pm0.007$		&	   $24^{+11}_{-10}$		&	$0.011^{+0.008}_{-0.006}$			&			\\

\ion{Ar}{\ K$\alpha$}	&	$4.197\pm0.005$		&		   $7^{+5}_{-4}$		&	0.005	$^{+0.003}_{-0.001}$		&	7	&	$4.189^{+0.008}_{-0.009}$		&	   $10^{+8}_{-6}$		&	0.007	$^{+0.001}_{-0.002}$		&	4		\\	
																												
\ion{S}{xvi} 												&&&&&			$4.729^{+0.171}_{-0.030}$		&	  $4^{+6}_{-4}$		&	0.005	$^{+0.005}_{-0.000}$		&	2		\\	

\ion{S}{xv $r$}	&$5.044^{+0.006}_{-0.005}$	   		& $2.8^{+4.1}_{-2.7}$			&0.006	$^{+0.002}_{-0.001}$		&	1.7 		&				\\														

\ion{S}{xv $f$}	&&&&&			$5.091^{+0.008}_{-0.012}$		&	   $7^{+8}_{-5}$		&	0.005$^{+0.003}_{-0.000}$&	4\\	
\ion{S}{\ K$\alpha$} 	&	$5.377\pm0.004$		&		  $16^{+7}_{-6}$		&	0.006	$^{+0.002}_{-0.001}$		&	14	&	$5.371\pm0.004$		&	   $40^{+15}_{-12}$		&	0.006	$^{+0.002}_{-0.002}$		&	27		\\	
																									
\ion{Si}{xiv} { Ly$\alpha$} 	&	$6.190^{+0.014}_{-0.011}$		&		   $9\pm4$		&	$0.026^{+0.019}_{-0.015}$			&	12	&	$6.196^{+0.020}_{-0.023}$		&	   $11^{+0.8}_{-0.5}$		&	$0.037^{+0.035}_{-0.018}$			&	9		\\	
																				
\ion{Si}{xiii $r$}	&	$6.648^{+0.001}_{-0.009}$		&		  $10\pm4$		&$0.023^{+0.007}_{-0.004}$			&	30	&	$6.644^{+0.006}_{-0.007}$		&	   $10^{+6}_{-5}$		&	$0.017^{+0.007}_{-0.005}$		&	23		\\	
																												
\ion{Si}{xiii $i$} 	&	$6.687^{+0.001}_{-0.009}$		&		  $2.4^{+3.0}_{-2.5}$		&$0.023^{+0.007}_{-0.004}$			&	30	& $6.683^{+0.006}_{-0.007}$			&	 $2.9^{+4.1}_{-2.9}$		&	$0.017^{+0.007}_{-0.005}$			&	23		\\	
\ion{Si}{xiii $f$}	&$6.741^{+0.001}_{-0.009}$			&		  $6.9^{+2.7}_{-2.3}$		&	$0.023^{+0.007}_{-0.004}$			&	30	&	$6.736^{+0.006}_{-0.007}$		&	  $10^{+5}_{-4}$	&	$0.017^{+0.007}_{-0.005}$		&	23		\\	
																					
\ion{Si}{xiv K$\alpha$} 	&	$7.119\pm0.005$		&		   $12.6^{+3.2}_{-2.8}$		&	$0.018^{+0.002}_{-0.004}$			&	20	&	$7.115\pm0.007$		&	  $25\pm0.6$		&	$0.026^{+0.004}_{-0.005}$			&	26		\\

\ion{Mg}{vxi} 	&	$7.212\pm0.007$		&		   $1.9^{+1.5}_{-1.1}$		&	$0.005^{+0.003}_{-0.000}$			&	1.9	&			&			&				&			\\	
%\hline

\ion{Al}{xii $r$}	&	$7.823^{+0.068}_{-0.023}$	&	  $1.1^{+1.5}_{-1.1}$	&	$0.007^{+0.000}_{-0.001}$	&	&	$7.816^{+0.075}_{-0.016}$	&	  $0.7^{+0.2}_{-0.7}$	&	$0.005^{+0.003}_{-0.001}$					\\
%\ion{Al} {xii}&		&		&		&	&	$7.800^{+0.299}_{-0.000}$	&	  $2.8^{+21.5}_{-2.8}$	&	$0.7^{+0.4}_{-0.7}$				&	\\
\ion{Mg}{xii} 	&	$8.455\pm0.013$	&	   $2.8^{+2.1}_{-1.5}$	&	$0.013^{+0.021}_{-0.010}$	&	&	$8.417^{+0.020}_{-0.017}$	&	   $11^{+6}_{-5}$	&	$0.040^{+0.034}_{-0.019}$				&	\\
\ion{Mg}{xi $r$}	&	$9.151^{+0.012}_{-0.011}$	&	   $3.9^{+3.4}_{-2.3}$	&	$0.007^{+0.000}_{-0.003}$	&	&	$9.115^{+0.011}_{-0.008}$	&	  $5^{+5}_{-4}$	&	$0.007^{+0.008}_{-0.003}$				&	\\
\ion{Mg}{xi $i$}	&	9.212317	&	   $2.4^{+3.1}_{-1.9}$	&	0.008	&	&	9.176157	&	  $3.1^{+4.5}_{-2.4}$	&	$0.007^{+0.008}_{-0.003}$				&	\\
\ion{Mg}{xi $f$}	&	9.296917	&	  $2.1^{+2.9}_{-2.0}$	&	0.008	&	&	9.260757	&	   $5^{+6}_{-4}$	&	$0.007^{+0.008}_{-0.003}$				&	\\

\ion{Ne}{x $\gamma$}	&	$9.799^{+0.201}_{-0.010}$	&	  $1.3^{+0.4}_{-1.3}$	&	$0.005^{+0.003}_{-0.005}$	&	&	$9.700^{+0.037}_{-0.024}$	&	   $5^{+8}_{-4}$	&	$0.014^{+0.058}_{-0.009}$				&	\\

%&&&&&&&&\\
\ion{Fe}{ \textsc{xx}}	&		&	  	&	&	&	$9.84^{+0.17}_{-0.05}$	&	  $2.9^{+5.2}_{-3.0}$	&	$0.005^{+0.003}_{-0.003}$				&	\\
%Fe \textsc{xx} &&&&&&&&\\
\ion{Ne}{x $\beta$}	&	$10.216^{+0.056}_{-0.017}$	&	   $7^{+6}_{-4}$	&	$0.07^{+0.06}_{-0.04}$	&	&	&	  	&					&	\\
%&&&&&&&&\\
\ion{Fe}{xxii} 	&	$11.16^{+0.09}_{-0.13}$	&	   $5^{+7}_{-4}$	&	$0.017^{+0.084}_{-0.012}$	&	&	$11.56^{+0.04}_{-0.37}$	&	  $6^{+12}_{-6}$	&	$0.007^{+0.000}_{-0.003}$				&	\\
%&&&&&&&&\\
\hline

\ion{Ne}{x $\alpha$ } 	&	$12.115^{+0.014}_{-0.015}$	&	   $6^{+9}_{-5}$	&	$0.005^{+0.003}_{-0.005}$	&	&	$12.126^{+0.012}_{-0.027}$	&	   $14^{+15}_{-11}$	&	$0.007^{+0.000}_{-0.003}$				&	\\
%&&&&&&&&\\
%&&&&&&&&\\

\ion{Ne}{ix $r$} 	&	$13.49^{+0.00}_{-0.09}$	&	   $11^{+15}_{-9}$	&	$0.007^{+0.003}_{-0.003}$	&	&	$13.487^{+0.014}_{-0.087}$	&	  $17^{+34}_{-17}$	&	$0.007^{+0.000}_{-0.003}$				&	\\
%&&&&&&&&\\

\ion{Ne}{ix $i$}	&	$13.52^{+}_{-}$	&	   $3^{+}_{-}$	&	$0.00014^{+0.00}_{-0.00}$	&	&	$13.56^{+}_{-}$	&	  0	&	$0.002^{+0.000}_{-0.00}$				&	\\
\ion{Ne}{ix $f$}	&	$13.66^{+0.00}_{-0.09}$	&	   $0$	&	$0.00014^{+0.00}_{-0.00}$	&	&	$13.705^{+0.0}_{-0.0}$	&	  	&	$0.02^{+0.000}_{-0.00}$				&	\\

\ion{Ne} {K$\alpha$} 	&	$14.79^{+0.00}_{-0.49}$	&	   $20^{+28}_{-19}$	&	$0.099^{+0.000}_{-0.094}$	&	&		&		&					&	\\

\hline\hline
\end{tabular}																		
																		
\end{threeparttable}	

%\end{center}																		
}																		
\end{table*}

%%%%New table 4u1700 quiescence flare out eclipse%%%%%
\begin{table*}
{\def\arraystretch{1.3}
%\begin{center}
\caption{Out-of-Eclipse emission lines (ObsID 17630).
\label{tab:Line_Emission_Out_Eclipse }}
\begin{threeparttable}
\label{tab:Line_Emission_Out_Eclipse}

\begin{tabular}{lcccccccc}
\hline
& & Quiescence& & &  & Flare & &\\																			
\hline

Line  &    $\lambda$             &  Line Flux        & $\sigma$   &Bayesian   & $\lambda$ &  Line Flux          & $\sigma$&Bayesian  \\
      &                          &  $\times 10^{-6}$  &          & Blocks     &           &    $\times 10^{-6}$ &         & Blocks   \\
      & (\AA)                    & (ph s$^{-1}$ cm$^{-2}$)  &(\AA) & $\alpha$$_\mathrm{sig}$     &  (\AA)&  (ph s$^{-1}$ cm$^{-2}$)& (\AA)      &  $\alpha$$_\mathrm{sig}$  \\
      
\hline

\ion{Fe}{\ K$\beta$}	&	1.753$^{+0.003}_{-0.004}$	&	190$^{+180}_{-90}$ 	&	0.005	&		&	1.756$^{+0.052}_{-0.052}$ 	&	251$^{+400}_{-150}$	&	0.005	&		\\

\ion{Fe}{xxvi  Ly$\alpha$} 	&	1.777$^{+0.002}_{-0.006}$	&	125$^{+80}_{-125}$	&	0.005	&		&	1.777$^{+0.002}_{-0.011}$	&	-240$^{+290}_{-430}$ 	&	0.005	&		\\

\ion{Fe }{xxv} 	&	1.855$^{+0.004}_{-0.000}$ 	&	170$^{+100}_{-50}$ 	&	0.005	&		&	1.855$^{+0.004}_{-0.004}$	&	170$^{+100}_{-50}$ 	&	0.005	&		\\

\ion{Fe}{ \ K$\alpha$} 	&	1.935$^{+0.003}_{-0.003}$ 	&	330 $^{+120}_{-90}$	&	0.005	&	4	&	1.939 $^{+0.004}_{-0.003}$	&	970$^{+30}_{-380}$	&	0.005	&	1.7	\\		
& & & & &  &  & &\\
																			
\ion{Ar}{xxvii} 	&	$3.363^{+0.017}_{-0.043}$	&	  $34^{+32}_{-30}$	&	0.005	&		&	$3.361^{+0.008}_{-0.009}$	&	   $220\pm120$	&	0.005	&		\\

\ion{Ar}{\ K$\alpha$ } 	&	$4.185^{+0.012}_{-0.005}$	&	   $17^{+18}_{-16}$	&	0.005	&		&	$4.180^{+0.021}_{-0.010}$	&	  $30^{+70}_{-40}$	&	0.005	&		\\		
																			
& & & & &  &  & &\\																			
\ion{S}{xv $r$}	&	$5.045^{+0.005}_{-0.043}$	&	  $3^{+12}_{-4}$	&	0.005	&		&	$5.036\pm0.008$	&	   $60^{+60}_{-50}$	&	0.005	&		\\

\ion{S}{xv $i$}	&	$5.071^{+0.005}_{-0.043}$	&	  $3^{+11}_{-4}$	&	0.005	&		&	$5.0622\pm0.008$	&	  $0$	&	0.005	&		\\

\ion{S}{xv $f$}	&	$5.108^{+0.005}_{-0.043}$	&	 $8^{+11}_{-8}$	&	0.005	&		&	$5.0989\pm0.008$	&	  $21^{+49}_{-21}$	&	0.005	&		\\		
& & & & &  &  & &\\																			
																			
\ion{S}{vi-vii  \ K$\alpha$}	&	$5.359^{+0.005}_{-0.006}$	&	   $20^{+15}_{-12}$	&	0.005	&		&	$5.323\pm0.006$	&	   $80^{+60}_{-50}$	&	0.005	&		\\		
& & & & &  &  & &\\																			
\ion{Si}{xiv ly$\alpha$}	&	$6.167^{+0.023}_{-0.042}$	&	   $2.7^{+3.5}_{-2.5}$	&	0.005	&	7	&	$6.190^{+0.016}_{-0.015}$	&	   $10^{+14}_{-10}$	&	0.005	&		\\

& & & & &  &  & &\\																			
\ion{Si}{xiii $r$}	&	$6.644^{+0.006}_{-0.025}$	&	  $4^{+5}_{-4}$	&	0.005	&	9	&	$6.647^{+0.004}_{-0.005}$	&	   $10^{+12}_{-8}$	&	0.005	&		\\		
																			
\ion{Si}{ xiii $i$}	&	$6.6831^{+0.006}_{-0.025}$	&	  $3\pm4$	&	0.005	&	9	&	$6.685^{+0.004}_{-0.005}$	&	   $11^{+12}_{-8}$	&	0.005	&		\\		
																			
\ion{Si}{xiii $f$}	&	$6.737^{+0.006}_{-0.025}$	&	 $2.4^{+3.1}_{-2.4}$	& 0.005		&	9	&	$ 6.738^{+0.004}_{-0.005}$	&	   $17^{+14}_{-10}$	&	0.005	&		\\		
& & & & &  &  & &\\																			
																			
\ion{Si}{  K$\alpha$}	&	$7.106\pm0.005$	&	   $11\pm5$	&	0.005	&	10	&	$7.109\pm0.007$	&	   $31^{+16}_{-13}$	&	0.005	&	3	\\		
																			
\ion{Mg}{xi}	&	$7.738^{+0.015}_{-0.013}$	&	   $2.3^{+2.7}_{-1.7}$	&	0.008	&		&	$7.84^{+0.06}_{-0.16}$	&	 $2.4^{+6.8}_{-2.5}$	&	0.005	&		\\		
																			
\hline\hline
\end{tabular}																		
\begin{tablenotes}
\item[a] Numbers without errors have been fixed at the quoted values. 

\end{tablenotes}																		
\end{threeparttable}	

%\end{center}																		
}																		
\end{table*}

\end{document}